\documentclass[aps,prl,twocolumn,superscriptaddress]{revtex4-1}

\usepackage{graphicx}
\usepackage{ragged2e}
\usepackage{amsmath}
\usepackage{xcolor}

\newcommand{\ket}[1]{|{#1}\rangle}

\begin{document} 

\title{Spin-charge separation in a 1D Fermi gas with tunable interactions} 

\author{Ruwan Senaratne}
\thanks{These authors contributed equally to this work.}
\affiliation{Department of Physics and Astronomy, Rice University, Houston, Texas 77005, USA}

\author{Danyel Cavazos-Cavazos}
\thanks{These authors contributed equally to this work.}
\affiliation{Department of Physics and Astronomy, Rice University, Houston, Texas 77005, USA}

\author{Sheng Wang}
\affiliation{State Key Laboratory of Magnetic Resonance and Atomic and Molecular Physics,Wuhan Institute of Physics and Mathematics, APM, Chinese Academy of Sciences, Wuhan 430071, China}
\affiliation{University of Chinese Academy of Sciences, Beijing 100049, China}

\author{Feng He}
\affiliation{State Key Laboratory of Magnetic Resonance and Atomic and Molecular Physics,Wuhan Institute of Physics and Mathematics, APM, Chinese Academy of Sciences, Wuhan 430071, China}
\affiliation{SISSA and INFN, Sezione di Trieste, 34136 Trieste, Italy}

\author{Ya-Ting Chang}
\affiliation{Department of Physics and Astronomy, Rice University, Houston, Texas 77005, USA}

\author{Aashish Kafle}
\affiliation{Department of Physics and Astronomy, Rice University, Houston, Texas 77005, USA}

\author{Han Pu}
\affiliation{Department of Physics and Astronomy, Rice University, Houston, Texas 77005, USA}

\author{Xi-Wen Guan}
\thanks{Corresponding author. Email: randy@rice.edu; xiwen.guan@anu.edu.au}
\affiliation{State Key Laboratory of Magnetic Resonance and Atomic and Molecular Physics,Wuhan Institute of Physics and Mathematics, APM, Chinese Academy of Sciences, Wuhan 430071, China}
\affiliation{Department of Theoretical Physics, RSPE,
Australian National University, Canberra ACT 0200, Australia}

\author{Randall G. Hulet}
\thanks{Corresponding author. Email: randy@rice.edu; xiwen.guan@anu.edu.au}
\affiliation{Department of Physics and Astronomy, Rice University, Houston, Texas 77005, USA}

\date{\today}
\begin{abstract}
Ultracold atoms confined to periodic potentials have proven to be a  powerful tool for quantum simulation of complex many-body systems.  We  confine fermions to one-dimension to realize the Tomonaga-Luttinger liquid model describing the highly collective nature of their low-energy excitations.  We use Bragg spectroscopy to directly excite either the spin or charge waves for various strength of repulsive interaction. We observe that the velocity of the spin and charge excitations shift in opposite directions with increasing interaction, a hallmark of spin-charge separation. The excitation spectra are in quantitative agreement with the exact solution of the Yang-Gaudin model and the Tomonaga-Luttinger liquid theory. Furthermore, we identify effects of non-linear corrections to this theory due to band curvature and back-scattering.
\end{abstract}

\maketitle
\small

Unlike three-dimensional (3D) metals whose low-energy excitations are fermionic quasi-particles, the low-energy excitations of one-dimensional (1D) fermions are collective bosonic spin- and charge-density waves (SDW/CDW) that disperse linearly, as described by the Tomonaga-Luttinger liquid (TLL) theory \cite{Tomonaga1950,Luttinger1963,Haldane1981,Voit1993,Giamarchi_Book}. Underlying this description is the remarkable result that the SDW and the CDW of an interacting 1D Fermi gas propagate at different speeds, thus causing a spatial separation of spin and charge excitations in the gas. 

Spin-charge separation has been studied in quasi-1D solid state materials in several ground-breaking experiments employing momentum-resolved tunneling to determine the dispersions \cite{Auslaender2002,Auslaender2005,Jompol2009} or by angle-resolved photoemission spectroscopy \cite{Segovia1999,Kim1996,Kim2006}. While these experiments observe splitting into spin and charge excitations, a quantitative analysis of these data has proved challenging because of the complexity of the electronic structure, and by the unavoidable presence of impurities and defects. Recently, a series of experiments with ultracold atoms in an optical lattice was performed on a single-site resolved 1D Hubbard chain, leading to the observation of the fractionalization of spin and charge quantum numbers at equilibrium \cite{Hilker2017}; the modification of the SDW wavevector by density-doping and by spin-polarization \cite{salomon2019}; and the study of simultaneous spin and charge dynamics outside the Luttinger liquid regime that result from a deconfinement-induced quench \cite{Vijayan2020}. These experiments, while demonstrating the abilities of ultracold atoms for quantum simulation with unprecedented control, did not measure the collective low-energy excitation spectrum inherent to spin-charge separation. The ability to make quantitative measurements of the respective speeds of the two low-energy modes, in a well-controlled setting with tunable interactions, has remained elusive.

The excitation spectrum of the charge (density) mode of fermionic atoms confined to quasi-1D tubes has been previously measured for fixed interaction \cite{Pagano2014}, and by us for variable interaction strength \cite{Ernie2018}.  These experiments used two-photon stimulated Bragg spectroscopy, illustrated in Figs. 1A and 1B, to impart an observable momentum $\hbar q$, with energy $\hbar\omega$, while the internal state of the atom is unchanged \cite{Stenger1999,Steinhauer2002,Veeravalli2008,Hoinka2012}. The response of the 1D gas at a particular $q$ and $\omega$ is related to the dynamic structure factor (DSF) $S(q, \omega)$, which characterizes the low-energy excitation spectrum for $q\ll k_F$, where $k_F$ is the Fermi wave vector. In our previous work the charge-mode structure factor $S_C(q, \omega)$ was measured and quantitatively compared with theory with good results \cite{Ernie2018}. Measurement of the spin-wave spectrum $S_S(q, \omega)$ remained out of reach because, without special measures, such a measurement induces single-photon scattering events that produce significant atom-loss. 

\begin{figure}[ht!]
\includegraphics[width= 0.85\linewidth]{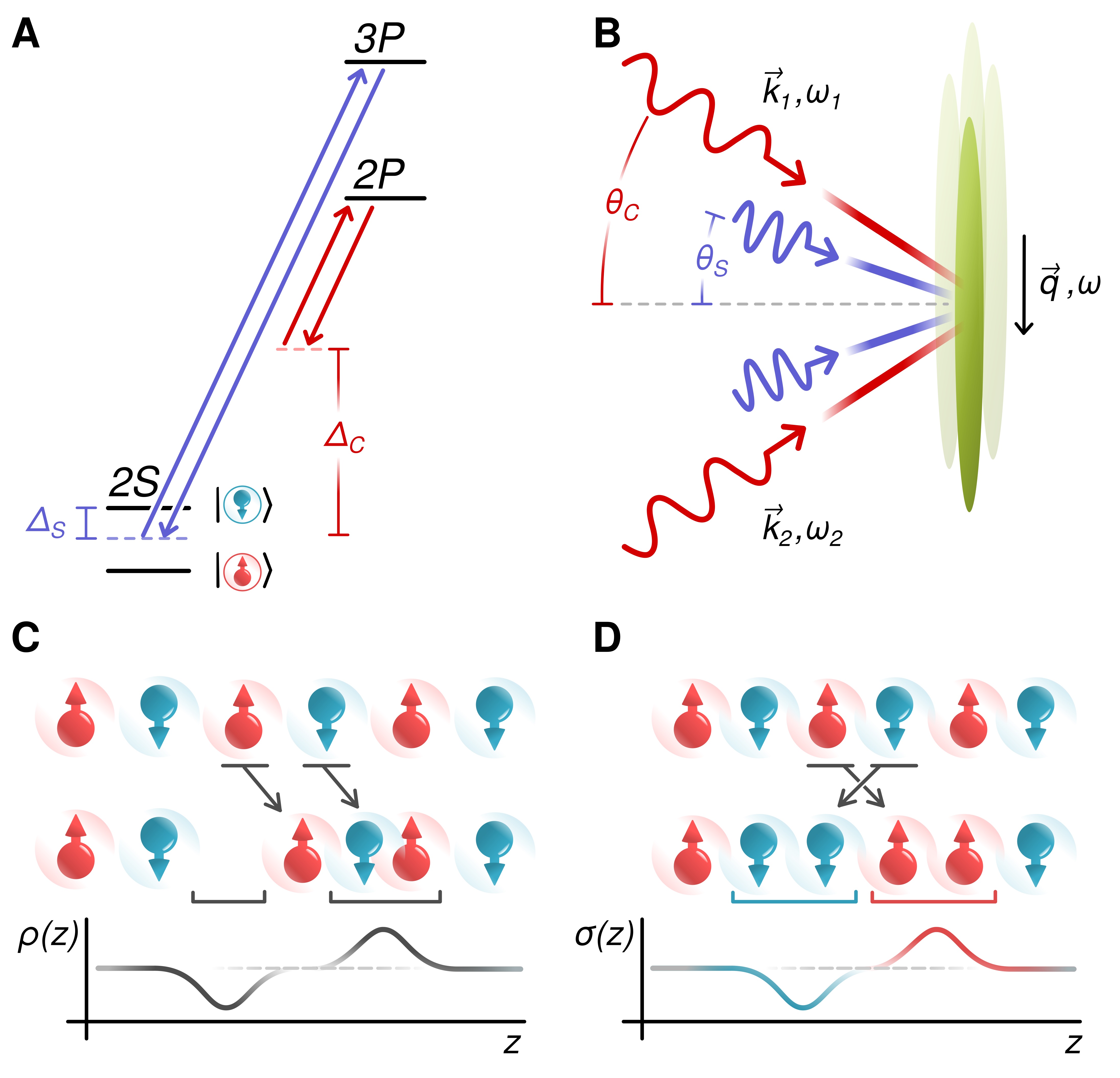}

\noindent \justifying{\textbf{Fig. 1. Spin and charge excitations via Bragg spectroscopy}. (\textbf{A}) Partial energy-level diagram of $^6$Li showing relevant
transitions and laser detunings for spin ($\Delta_S$, violet) and charge ($\Delta_C$, red) excitations. (\textbf{B}) Relative orientation ($\theta_{C,S}$) of each Bragg beam (1, 2) with respect to the axis perpendicular to the 1D tube direction .  A momentum transfer $\vec{q} = \vec{k}_1 - \vec{k}_2 \approx 0.2 \; k_F $  for the central tubes is delivered to the sample for a given relative detuning $\omega = \omega_1 - \omega_2$ between the beams. (\textbf{C} and \textbf{D}) Schematic diagram of the charge and spin excitations, showing an excitation of a particle-hole pair, or a spinon pair, respectively. The effect on the total density $\rho(x)$ and spin density $\sigma(x)$ is shown for each case at the bottom. The excitations are depicted, for clarity, as starting from a classical zero-temperature antiferromagnetic ground state in the strongly repulsive regime.}
\end{figure}

We have developed novel improvements to our implementation of Bragg spectroscopy to reduce spontaneous scattering to an acceptable level while, for the first time, selectively exciting either the SDW or the CDW with tunable repulsive interactions. The measurements are compared to the TLL theory, which describes the low-energy excitations of the more general Yang-Gaudin model of a spin-1/2, 1D Fermi gas with repulsive delta-function interactions in the continuum limit \cite{Yang1967,Gaudin1967}. They provide a unique quantitative test of spin-charge separation in the Luttinger liquid regime. Additionally we quantitatively show that our results for $S_S(q,\omega)$ provide evidence for deviations from the linear TLL theory due to low-energy back-scattering interactions, which are usually neglected to obtain a linear spin-mode dispersion \cite{Giamarchi_Book}.

Bragg spectroscopy is well suited to studying spin-charge separation, as the CDW or SDW may be isolated by the choice of detuning of the Bragg beams from resonance with an electronic excited state (Fig. 1). The detuning determines the sign of the light shift potential, which can be used to create a symmetrical light shift that exclusively excites charge waves, or an asymmetrical potential that only excites spin waves. For a system composed of a balanced mixture of two spin components ($\uparrow$,$\downarrow$) we can identify two independent contributions to the DSF, $S_{\uparrow\uparrow}$ and $S_{\uparrow\downarrow}$, and thus define a charge- and spin-density DSF given by \cite{Giamarchi_Book}

\begin{equation}
S_{C,S}(q,\omega) \equiv 2\left[S_{\uparrow\uparrow}(q,\omega) \pm S_{\uparrow\downarrow}(q,\omega) \right],
\label{eq:s}
\end{equation}

\noindent  where the $+$ sign corresponds to charge and $-$ to spin. At zero temperature, the momentum transfer to the system from the Bragg beams is given by \cite{Pines_book,Hoinka2012} 

\begin{equation}
P(q,\omega) \propto \left(\frac{1}{\Delta^2_\uparrow}+\frac{1}{\Delta^2_\downarrow}\right)S_{\uparrow\uparrow}+\frac{2}{\Delta_\uparrow\Delta_\downarrow}S_{\uparrow\downarrow},
\label{eq:pD_}
\end{equation}

\noindent where $\Delta_\sigma$ is the relative detuning of the Bragg beam from the excited state with respect to each ground spin state $\sigma$. If the conditon $\Delta_{\uparrow} \approx \Delta_{\downarrow} \gg \Delta_{\uparrow\downarrow}$ is satisfied, where $\Delta_{\uparrow\downarrow}$ is the splitting of the spin states, then $P(q,\omega) \propto S_C(q,\omega)$ and a CDW is excited, as depicted in Fig. 1C. On the other hand, if $\Delta_{\uparrow} = - \Delta_{\downarrow} = |\Delta_{\uparrow\downarrow}|/2$, then $P(q,\omega) \propto S_S(q,\omega)$ and an SDW is excited, as depicted in Fig. 1D. The detuning required for measuring $S_S(q,\omega)$ is thus fixed by $\Delta_{\uparrow\downarrow}$, unlike in the case for measuring $S_C(q,\omega)$, where in principle the detuning may be arbitrarily large. In the finite temperature case, a reverse Bragg process must also be considered, for which the momentum transfer is modified accordingly as $P(q,\omega) \propto S(q, \omega)-S(-q, -\omega)=S(q, \omega)(1-\mathrm{exp}(-\hbar\omega/k_\mathrm{B} T))$ \cite{Brunello2001,Cherny2006}.

In order to reduce spontaneous scattering during the Bragg measurement, the ratio of $\Delta_{\uparrow\downarrow}$ to $\Gamma$, the linewidth of the transition, must be increased. We approximately doubled $\Delta_{\uparrow\downarrow}$ by choosing $\ket{1}$ and $\ket{3}$ as our pseudo-spin-1/2 states, rather than $\ket{1}$ and $\ket{2}$ as used previously (states $\ket{1},\,\ket{2}$ and $\ket{3}$ are the three lowest hyper-fine states of $^6$Li) \cite{Ernie2018,Veeravalli2008}. For the excitation of the SDW, we took the additional step of reducing $\Gamma$ by detuning the Bragg beams from the $3P_{3/2}$ excited state at a wavelength of 323 nm, rather than the usual 671 nm transition to the $2P_{3/2}$ state we use to excite the CDW (see Fig. 1A). The spontaneous-decay linewidth of the ultraviolet transition is nearly 8 times smaller than that for the red transition \cite{Duarte2011}. We compensate the difference in wavelength by simply adjusting the angle between the Bragg beams ($\theta_c \simeq 4.5^\circ$ and $\theta_s \simeq 2.2^\circ$, see Fig. 1B), such that for both cases the Bragg wave-vector is parallel to the tube axis and has a magnitude $|\vec{q}|=1.47$ $\mu$m$^{-1}$, corresponding to $0.2\,k_F$ for a peak-occupancy tube. Thus, the only net effect of this change is to further reduce the rate of incoherent scattering. The combination of these two steps reduces the spontaneous scattering by more than a ten-fold factor for a given Bragg coupling, as compared with Ref. \cite{Ernie2018}, and is sufficient to measure $S_S(q, \omega)$.

A more detailed description of our experimental methods may be found in the supplementary materials \cite{SM}.  We prepare a spin-balanced mixture of $^6$Li atoms in the two energetically lowest hyperfine sublevels, states $\ket{1}$ and $\ket{2}$, and confine them in an isotropic optical trap. We evaporatively cool the atoms to a temperature $T\approx 0.1\,T_F$, where $T_F$  is the Fermi temperature. We create an effectively 1D system  which realizes the Yang-Gaudin model, by loading the atoms into a 2D optical lattice with depth of $15\,E_r$, the recoil energy of a lattice photon of wavelength 1.064 $\mu$m. The resulting trap configuration is an array of quasi-1D tubes that are elongated in the axial dimension with an aspect ratio of $\sim$170. 

The number of atoms per tube is non-uniform across the ensemble of tubes due to the Gaussian curvature of the optical beams. The number distribution from tube-to-tube is also dependent on interaction strength; the central tube occupancy is highest for a non-interacting gas and decreases as repulsive interactions increase. We actively control for these variations by applying a focused repulsive green (532 nm) laser beam during the lattice ramp-up along each of the three orthogonal axes \cite{Hart15}. Varying the depth of this harmonic anti-trapping potential allows us to adjust the amount of confinement produced by the optical lattice, and thus make the density profiles comparable between different interaction strengths while keeping the total atom number constant. We measure the tube occupancy by taking \textit{in situ} phase-contrast images of the atom cloud \cite{Randy2020} and performing an inverse Abel transform to obtain the 3D distribution. A typical ensemble consists of a total of $6.5\times10^4$ atoms, has a peak tube occupancy of ${\sim}50$ atoms and a most probable tube occupancy of ${\sim}30$ atoms.

We perform Bragg spectroscopy by applying the pair of Bragg beams on the atoms in a 200 $\mu$s pulse. The intensity per beam is fixed to limit the loss of atoms due to spontaneous scattering to 6-8\% during the spin-mode measurement and to ensure that the momentum transfer is in the linear response regime for either mode over the entire range of interaction strengths we study \cite{SM}.  There is no discernible atom loss during the charge-mode measurement. Immediately after the Bragg pulse, the atoms are released from the lattice, and are imaged using phase-contrast imaging following 150 $\mu$s of time-of-flight, after which atoms receiving the Bragg kick are noticeably displaced from the center of the cloud (see Fig. S1 of the supplementary materials). We define the Bragg signal to be proportional to the number of out-coupled atoms. 

\begin{figure}[ht!]
\centering\includegraphics[width= 0.9\linewidth]{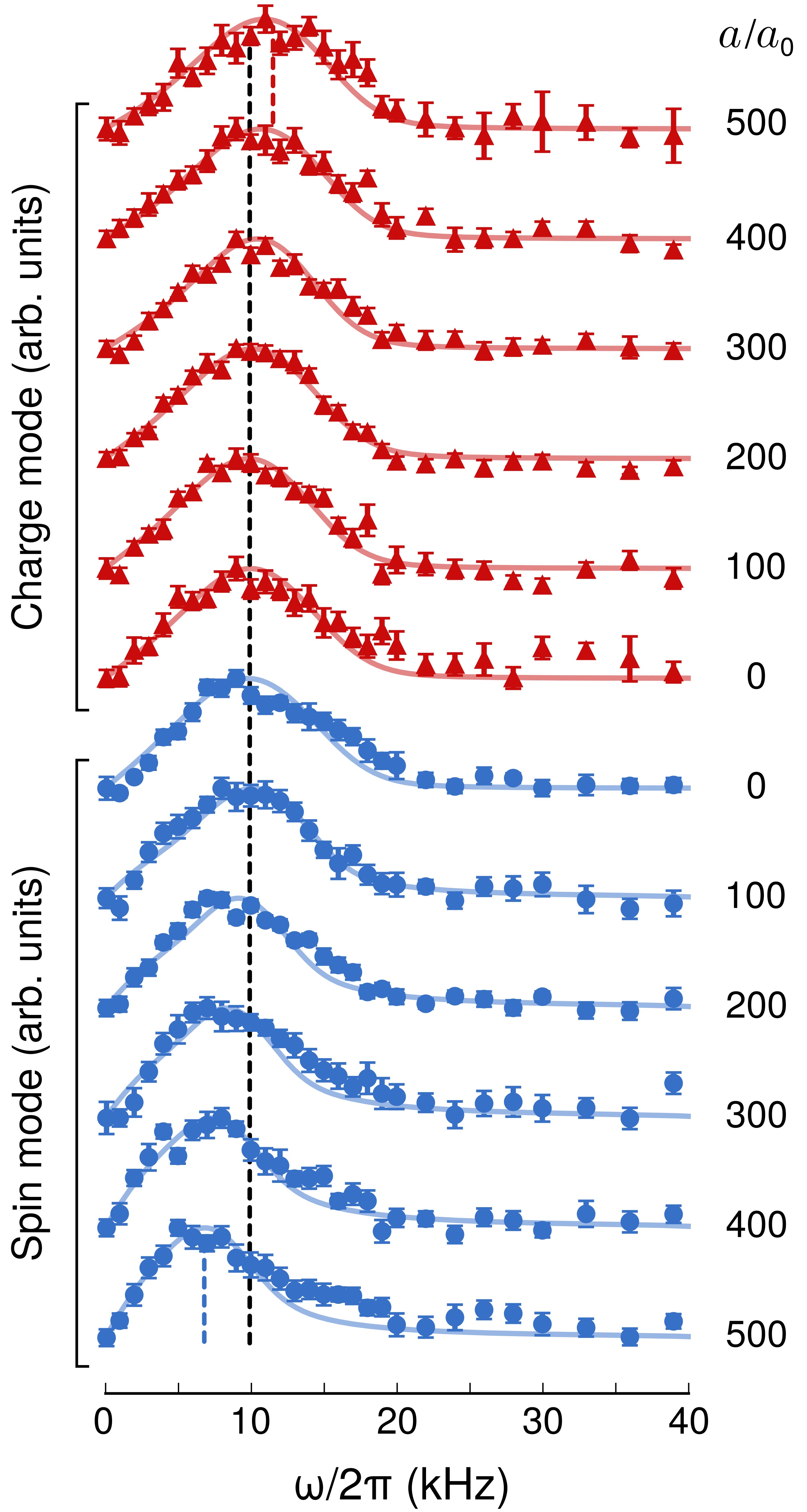}

\noindent \justifying{\textbf{Fig. 2. Bragg spectra}. Normalized Bragg signals related to $S_C(q,\omega)$ (red triangles) and $S_S(q,\omega)$ (blue circles) for the range of 3D scattering length $a$ from 0 to 500 $a_0$. Each data-point is the average of at least 20 separate experimental shots. Error bars represent standard error, obtained via bootstrapping \cite{Boot}. Vertical dashed lines show the extracted peak frequency $\omega_p$ for the non-interacting case (black), and the strongest probed interactions for the spin- and the charge-mode (blue and red, respectively). Solid lines are the calculated Bragg spectra for a global temperature $T$ = 250 nK with no additional fitting parameters other than overall scaling. Theory includes the non-linear effects of band curvature in the charge-mode and back-scattering in the spin-mode (for linear theory, see Figs. S7 and S8 of the supplemental materials). Deviations from theory at high frequency may be due to unaccounted-for corrections of order $q^3$.}
\end{figure}

The interaction strength is readily tunable using the Feshbach resonance between states $\ket{1}$ and $\ket{3}$ located at 690 G \cite{Zurn13}. The 3D scattering length $a$ may be tuned between $a$=0 and $a$ = 500 $a_0$, where $a_0$ is the Bohr radius, without appreciable loss. Figure 2 shows the measured (symbols) and calculated (solid lines) Bragg spectra for both modes in the range of $a$ from 0 to 500 $a_0$. Our DSF calculations take into account the effect of the inhomogeneous density due to the harmonic confinement along each tube by use of the local density approximation (LDA). The strength of interactions is density dependent and is given by the dimensionless Lieb-Liniger parameter $\gamma = {m g_1 (a)}/{\hbar^2 \rho_\mathrm{1D}}$, where $g_1(a)$ is the coupling strength of the quasi-1D pseudopotential \cite{Olshanii1998}, and $m$ is the atomic mass. The local density $\rho_\mathrm{1D}$ determines the local Fermi velocity and momentum ($v_F$ and $ \hbar k_F$), the Luttinger parameters ($K_{c,s}$), as well as the local velocities of the charge and spin waves ($v_c$ and $v_s$ respectively). As the Bragg signal is proportional to the total transferred momentum, we sum up the local values of the DSF along each tube, by invoking the LDA, to obtain the calculated spectra. Finally, we account for the frequency-broadening due to the finite duration of the Bragg pulses. A global temperature of $250$ nK is the only free parameter in this model, other than an independent normalization of each calculated and measured spectrum. 

In order to calculate the charge and spin DSFs, we use the exact Bethe ansatz solution of the Yang-Gaudin model at zero temperature \cite{Yang1967,He2020}. For $|q| \ll k_F$, the low-energy charge and spin excitations have approximately linear dispersion, and in this approximation $S_{C,S}(q,\omega)\propto |q|\,\delta (\omega-v_{c,s}q)$. However, at finite temperature and $q$ these DSFs are broadened when non-linear effects are considered \cite{Imambekov2012}. For the strength of interactions probed in this experiment, and for $T\ll T_F$, $S_C(q,\omega)$ is well approximated by the non-interacting DSF because the latter also exhibits a particle-hole excitation spectrum with width $\propto q^2$. As in previous work \cite{Ernie2018}, the effect of interactions is accounted for by replacing $k_F$ with $k_c=m^*v_c/\hbar$, where $m^*$ is the effective mass. The leading correction to the spin-mode dispersion at finite temperature is due to a low-energy back-scattering process which is expected for contact interactions in 1D \cite{Haldane1979}. Here, distinguishable spins permute between the two Fermi points by exchanging $2 k_F$. This process is exclusive to the spin sector and disrupts the linearization of the spin dispersion in the bosonization approach of TLL theory \cite{Giamarchi_Book,He2020}. We obtain the retarded spin-spin correlation function at finite temperature via the dressed spin-boson propagator \cite{Pereira2010}. By comparing our measurements to the non-linear Luttinger liquid (NLL) theory, we find that accounting for non-linearities due to back-scattering is necessary to model the spin Bragg spectra, particularly for large interactions. The linear TLL model fails to reproduce the observed high-frequency tails of the spin-mode Bragg spectra \cite{SM}.

The frequency at which the Bragg signal reaches a maximum, $\omega_\mathrm{p}$, corresponds to the most probable value of the mode velocity, $v_\mathrm{p}  = \omega_\mathrm{p}/q$, in the ensemble. We determine the peaks of each of the measured spectra by fitting a parabola to the data-points that are above $50\%$ of the maximum measured value for each spectrum. The locations of the peaks of the spectra obtained for our range of interaction are shown in Fig. 3, along with the peaks of the calculated spectra for each mode, which are in excellent agreement. For the non-interacting gas, the spin and charge collective modes have the same speed, resulting in nearly identical measured spectra for the two cases \cite{SM}. The congruence between the two spectra also confirms that the atom loss suffered during the spin-mode measurement has no discernible effect on the measured Bragg spectrum. As the strength of the interaction is increased, the charge-mode velocity $v_c$ increases, whereas the spin-mode velocity $v_s$ decreases. This is seen in the shifts of the peaks of the two spectra: to a lower frequency for the spin-mode, and to a higher frequency for the charge-mode.

\begin{figure}[ht!]
\centering\includegraphics[width= 0.9\linewidth]{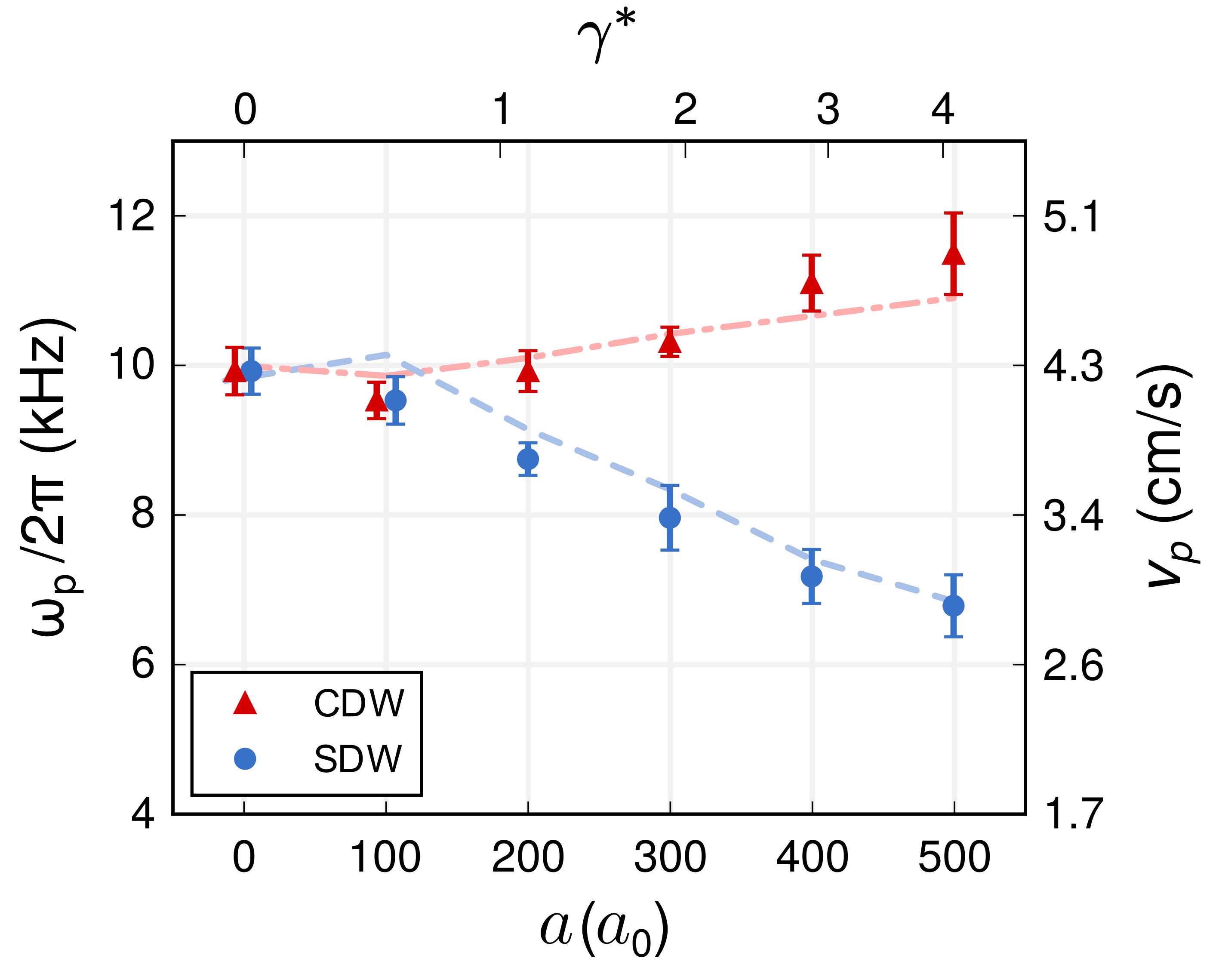}

\noindent \justifying{\textbf{Fig. 3. Spin-charge separation}. Peaks of measured Bragg spectra for charge (red triangles) and spin (blue circles) configurations for $a$ ranging from 0 to 500 $a_0$. Peak frequency values determined via fits of a parabolic function to the data-points above $50\%$ of the maximum measured value, and error bars are statistical standard errors of the relevant fit parameters. The corresponding speed of sound $v_p=\omega_p/q$ is given by the right axis. The upper horizontal axis gives the interaction strength in terms of the Lieb-Liniger parameter $\gamma^*$, evaluated at the center of a tube with occupancy 30 atoms. Lines show the calculated values for $\omega_p$ for the charge- and the spin-mode (dash-dotted red and dashed blue, respectively). Symbols for $a$ = 0 and 100 $a_0$ have been slightly displaced from one another for clarity.  Non-monotonicity in the charge-mode data and theory at low interaction is due to small residual differences in the number profiles prepared at different interaction strengths. Non-monotonicity in the spin-mode theory is likely due to neglecting the effects of band curvature, a $q^3$ correction.}
\end{figure}

We further explored the NLL regime by extracting the axial width of the out-coupled atom packet after time-of-flight expansion, as shown in Fig. 4 as functions of interaction for both modes. As expected, the out-coupled widths increase with $\gamma$ for measurements of the spin-mode, while remaining approximately constant for the charge-mode. We are able to model the increase in the out-coupled width for the spin-mode by calculating the spread in velocities implied by the finite spin-boson lifetime due to back-scattering \cite{SM}.

Having harnessed the tunability of interactions available in the cold-atom setting, we reveal the role of interactions in spin-charge separation for the first time, by tuning between a spin-charge separated regime and one where there is no separation. Further, the selectivity of the Bragg process in exciting either the CDW or the SDW allows us to provide the clearest demonstration of the division of the TLL Hamiltonian into distinct spin and charge sectors. Bragg spectroscopy may be used to probe the ultracold-atom TLL beyond the demonstration of spin-charge separation contained in this work. Measurements with variable $q$ can be conducted to further study the NLL and to benchmark novel calculations which include effects of band curvature and spin-charge coupling \cite{Imambekov2009,Pereira2010,Imambekov2012}. Additionally, at elevated temperatures and interactions, a spin-incoherent Luttinger liquid is expected, which supports a propagating charge-mode but not a spin-mode \cite{Fiete2007,Kakashvili2008}. Spin-imbalanced mixtures and attractive interactions are also of interest and are accessible via this technique \cite{Guan2013}. Experiments with shallower lattices will allow for the study of dimensionality effects due to tunneling between tubes \cite{Giamarchi_Book}. It is increasingly clear that the oft-admired mathematical elegance of 1D many-body physics is well complemented by the purity and tunability of ultracold atoms.

\begin{figure}[ht]
\centering\includegraphics[width= 0.85\linewidth]{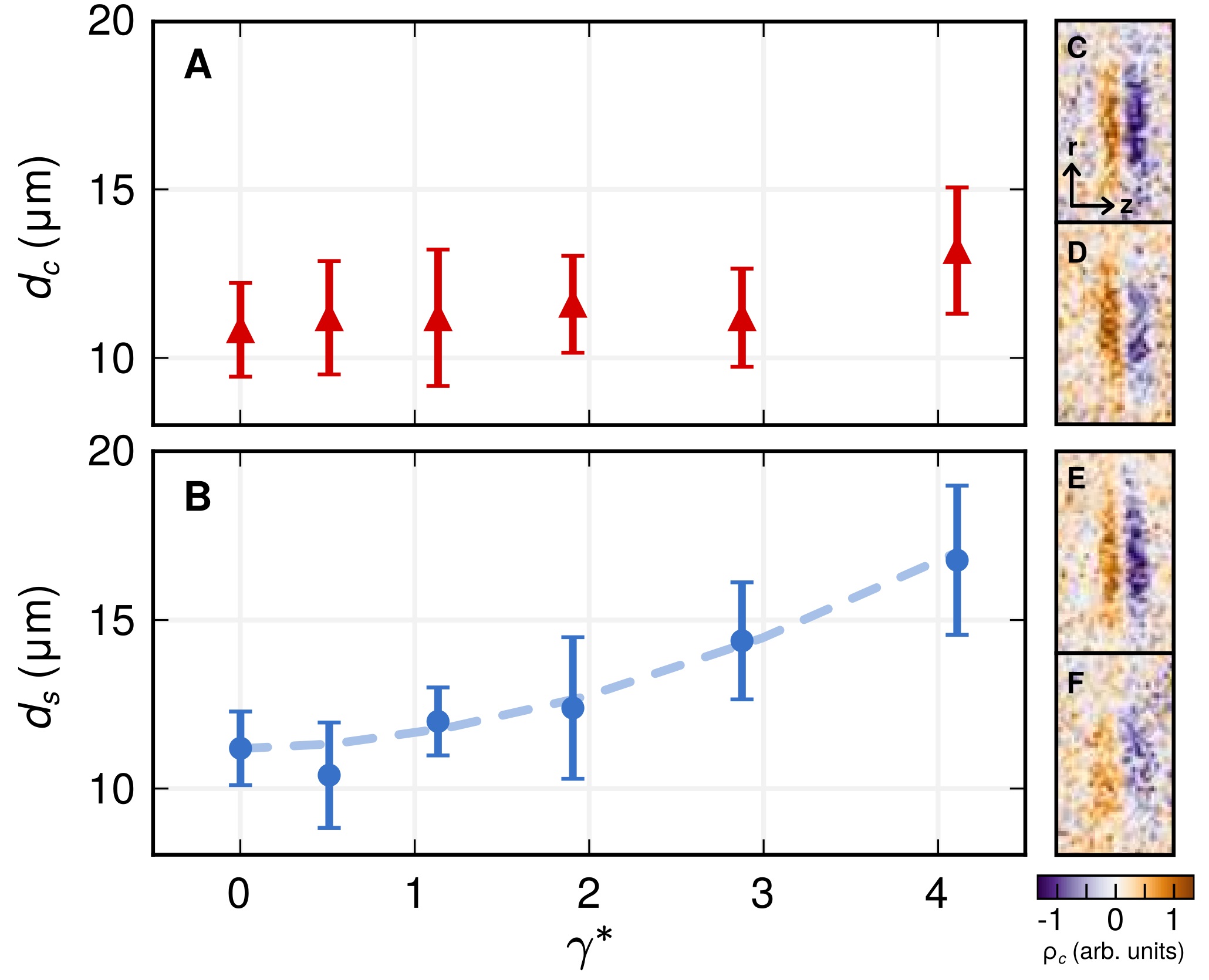}

\noindent \justifying{\textbf{Fig. 4. Dispersion of spin and charge density waves}. $1/e^2$ axial width of outcoupled atoms following a Bragg pulse and 150 $\mu$s time-of-flight for (\textbf{A}) charge ($d_c$, red triangles) and (\textbf{B}) spin ($d_s$, blue circles) excitations, with $a$ ranging from 0 to 500 $a_0$. The widths are the Gaussian fits to the postive outcoupled signal at $\omega_p$. Error bars are standard errors determined by bootstrapping for at least 20 independent images \cite{Boot}. The horizontal axis gives the Lieb-Liniger parameter $\gamma^*$ calculated for a median tube occupancy of 30 atoms \cite{SM}. The blue dashed line shows an estimation for $d_s$ derived from the finite lifetime of the spin bosons. (\textbf{C}-\textbf{F}) Representative samples of column-density ($\rho_c$) images of the atom cloud after the Bragg pulse and time-of-flight. (\textbf{C}) Charge mode with $a$ = 0 and (\textbf{D}) $a$ = 500 $a_0$. (\textbf{E},\textbf{F}) Spin mode with  $a$ = 0 and $a$ = 500 $a_0$, respectively. Each frame corresponds to 40 $\mu$m x 65 $\mu$m. }
\end{figure}

\bibliographystyle{Science}

\section*{Acknowledgments}
We would like to thank T. L. Yang for his contributions to
the apparatus. \textbf{Funding:} This work was supported in part by the
Army Research Office Multidisciplinary University
Research Initiative (Grant No. W911NF-17-1- 0323), the NSF (Grant Nos. PHY-1707992 and PHY-2011829), and the Welch Foundation (Grant No. C-1133). H. P. acknowledges support from the U.S. NSF (PHY-1912068) and the Welch Foundation (Grant No. C-1669). X.W.G. is supported by the NSFC key grant No. 12134015, the NSFC grant  No. 11874393, and the NKRDPC 2017YFA0304500. D.C.-C. acknowledges financial support from CONACyT (Mexico, Scholarship No. 472271). \textbf{Author contributions:} R.S and D C.-C. built the experimental setup, performed the measurements, and analyzed data.  S.W. and F.H performed the modeling of the experimental data and the theoretical calculations. Y.C. and A.K. assisted with experiments. All work was supervised by H.P, X.W.G. and R.G.H. All authors discussed the results and contributed to the manuscript. \textbf{Competing interests:} The authors declare no competing interests. \textbf{Data and materials availability:} All data needed to evaluate the conclusions in the paper are present in the paper and the supplementary materials.

\bibliography{UVSC}

\section*{Supplementary Materials}
\renewcommand{\theequation}{S\arabic{equation}}
\setcounter{equation}{0}

\subsection*{Materials and Methods}
The apparatus and experimental procedures used to produce degenerate Fermi gases for this study have been discussed previously \cite{Hart15,Ernie2018,Randy2020}. The primary differences with our previous experiment \cite{Ernie2018} are aimed at reducing spontaneous light scattering in the implementation of Bragg spectroscopy of the SDW. The sample is initially prepared in a spin-balanced mixture of the two lowest hyperfine sub-levels, which we label $\ket{1}$ and $\ket{2}$, respectively. At this stage, we trap $6.5\times10^4$ atoms in an isotropic harmonic trap with a geometric-mean trapping frequency of $(2\pi) \times$ 258 Hz produced by the intersection of three mutually-orthogonal focused trapping beams of wavelength $1.064$ $\mu$m. Each beam is linearly polarized and initially retro-reflected with a perpendicular linear polarization in order to avoid lattice formation. The sample temperature is $0.1\,T_F$, where $T_F$ is the Fermi temperature. We then transfer the state $\ket{2}$ atoms to the third lowest hyperfine sub-level $\ket{F= 3/2, m_F = -3/2}$, which we label $\ket{3}$, by applying a 110 $\mu$s $\pi$-pulse on the radio frequency magnetic dipole transition from $\ket{2}\rightarrow\ket{3}$. We tune the $\ket{1}-\ket{3}$ 3D $s$-wave scattering length, $a$, using a magnetic Feshbach resonance located at 690 G \cite{Zurn13}. The coupling strength of the quasi-1D pseudopotential  is given by $g_1(a) = -2 \hbar^2/m a_{\rm 1D}$, and the  effective 1D scattering length is given by $a_{\mathrm{1D}}=\left( -a_{\perp }^{2}/a\right) \left[ 1-C\left(
a/a_{\perp }\right) \right]$, where $C = |\zeta(1/2)|/\sqrt{2}\sim1.03$ \cite{Olshanii1998} and $a_\perp=\sqrt{\hbar/m\omega_r}$ is the length-scale of the transverse harmonic confinement with frequency $\omega_r$. We perform the state transfer at the magnetic field corresponding to the desired value of $a$ for each experimental run. The magnetic field is stabilized to $\pm 10$ mG using a two-stage servo\cite{YaTing2020}.

We then rotate the polarization of the retro-reflected beams by 90 degrees in order to form a shallow 3D lattice, and afterwards ramp up the intensities of the three trapping beams within 35 ms to reach a depth of 7 $E_r$ along each axis, where $E_r = h\times 29.4$ kHz is the recoil energy of a single trapping photon. We simultaneously turn on focused, but not retroreflected, blue-detuned anti-trapping beams (532 nm) along each of the lattice axes. These `compensation' beams are used to adjust the potential at the center of the trap, in order to achieve a consistent number profile across the ultimate ensemble of quasi-1D tubes for all studied interaction strengths\cite{Ernie2018,Hart15}. 

Having loaded the atoms into the compensated 3D lattice, we ramp up the intensity of two of the trapping beams to reach a trapping depth of 15 $E_r$ along two of the lattice axes in 20 ms, while simultaneously ramping the third lattice beam and compensation beam intensities to zero. The resultant 2D lattice forms an array of quasi-1D tubes. Each tube has a radial angular trapping frequency $\omega_r = 2\pi\times227.5$ kHz and axial angular trapping frequency $\omega_z = 2\pi\times1.34$ kHz, giving a tube aspect ratio ${\sim}170$. We load an on-average spin-balanced sample of up to 60 atoms in each tube, with a temperature ${\sim}250$ nK. Thus, the chemical potential $\mu \ll \hbar \omega_r$, and we avoid populating radially excited modes. The energy scale associated with tunneling between neighboring tubes is ${\sim} h\times200$
Hz, and is much smaller than the thermal energy-scale, ensuring that the ensemble is 1D in nature \cite{Giamarchi_Book}.

Once the 2D lattice has been ramped up, we apply the Bragg pulse, by turning on the two Bragg beams for 200 $\mu$s. We chose this time to minimize the pulse-time broadening while keeping the probe duration smaller than half the axial oscillation period. For the charge mode, we detuned the Bragg beams $\sim$11.4 GHz blue of the  $2S_{1/2}\rightarrow 2P_{3/2}$ ($D_2$) $\sigma^+$ transition. This relatively large detuning ensures we satisfy the CDW condition described by Eq.(\ref{eq:pD_}) in the main text, and only an effectively pure charge-wave is excited. Each Bragg beam has a $1/e^2$ radius (waist) of 500 $\mu$m and a power of 200 $\mu$W. 

For the spin-mode, the detunings of the Bragg beams from the $\pi$-transition to the $3P_{3/2}$ manifold are of opposite sign for the two spin states to satisfy the SDW condition, and equal in magnitude to half of the $\sim$159 MHz $\ket{1}-\ket{3}$ state-splitting. Each Bragg beam has a waist of 100 $\mu$m and a power of 168 $\mu$W. We adjust the angle between the Bragg beams to keep the magnitude of the momentum transfer the same for both transitions according to the Bragg relation $q = 4\pi\sin(\theta)/\lambda$, where $\lambda$ is the wavelength of the transition and $\theta$ the half-angle between the Bragg beams. The applied intensity $I$ is significantly higher in the case of the spin-mode Bragg beams due to the poor UV transmission of the vacuum window used for this experiment, which we cannot directly measure but indirectly estimate to be ${\sim}10\%$ via scattering rate measurements. 

At the end of the pulse, we turn off the lattice beams, and allow the atoms to expand freely for a further 150 $\mu$s, before imaging the cloud using polarization phase-contrast imaging (PPCI) \cite{Randy2020}. We take images for varying values of the frequency difference, $\omega$, between the Bragg beams. As shown in Fig. S1, an image of the outcoupled atoms is obtained by subtracting a reference image for which $\omega=0$, averaging over more than 20 independent experimental runs for both images. We sum the obtained difference image in the radial dimension, to obtain an axial line density which shows an excess of outcoupled atoms and a deficit at their origin. We subtract off any offset (which we determine by fitting), and then sum the absolute value of the line density to obtain the Bragg signal.

\begin{figure}[hb!]
\centering\includegraphics[width= 1.0\linewidth]{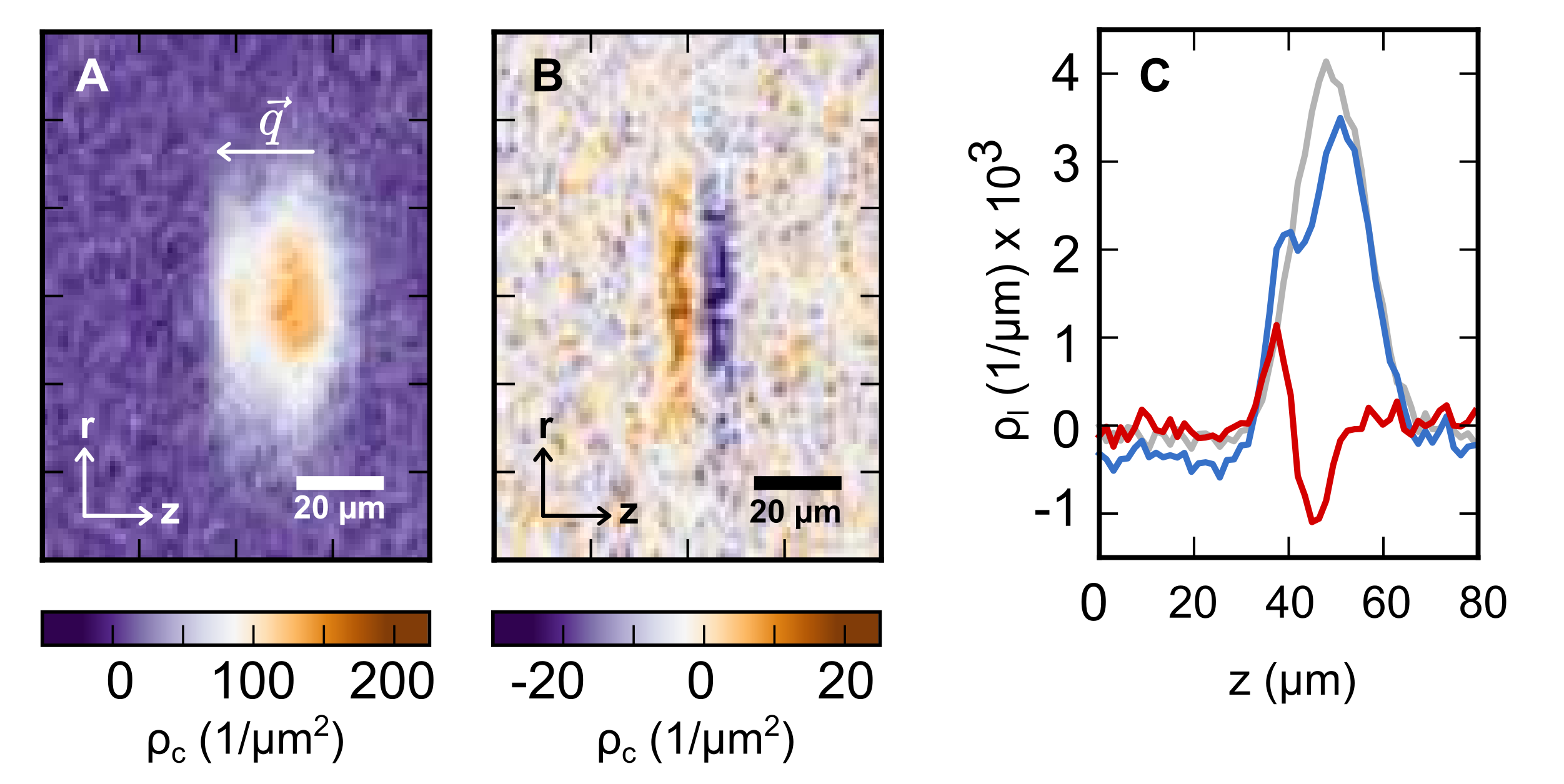}
\end{figure}
\noindent {\textbf{Fig. S1. Bragg signal extraction.} (\textbf{A}) Column density ($\rho_c$) image of the atom cloud after a Bragg pulse delivers a momentum transfer $\vec{q}$ to the sample, and the sample is released from the trap and allowed to expand during a  150 $\mu$s time-of-flight. (\textbf{B}) Difference between (\textbf{A}) and a reference shot with no relative detuning between the Bragg beams. (\textbf{C}) Line density ($\rho_l$) of (\textbf{A}) and (\textbf{B}) (blue and red, respectively), and the reference shot (gray). Line densities are calculated by integrating the column density along $r$, the axis perpendicular to the 1D tube direction $z$. The Bragg signal is calculated as the sum of the absolute value of the offset-subtracted line densities.}

For both modes we ensure that we are in the linear-response regime by varying $I$ and checking that the measured Bragg signal size is proportional to $I^2$. Beyond a certain intensity, we observe that the Bragg signal saturates due to depletion, as shown below in Fig. S2. We adjusted the probing power to remain below this limit for both cases.

\begin{figure}[ht!]
\centering\includegraphics[width= 1.0\linewidth]{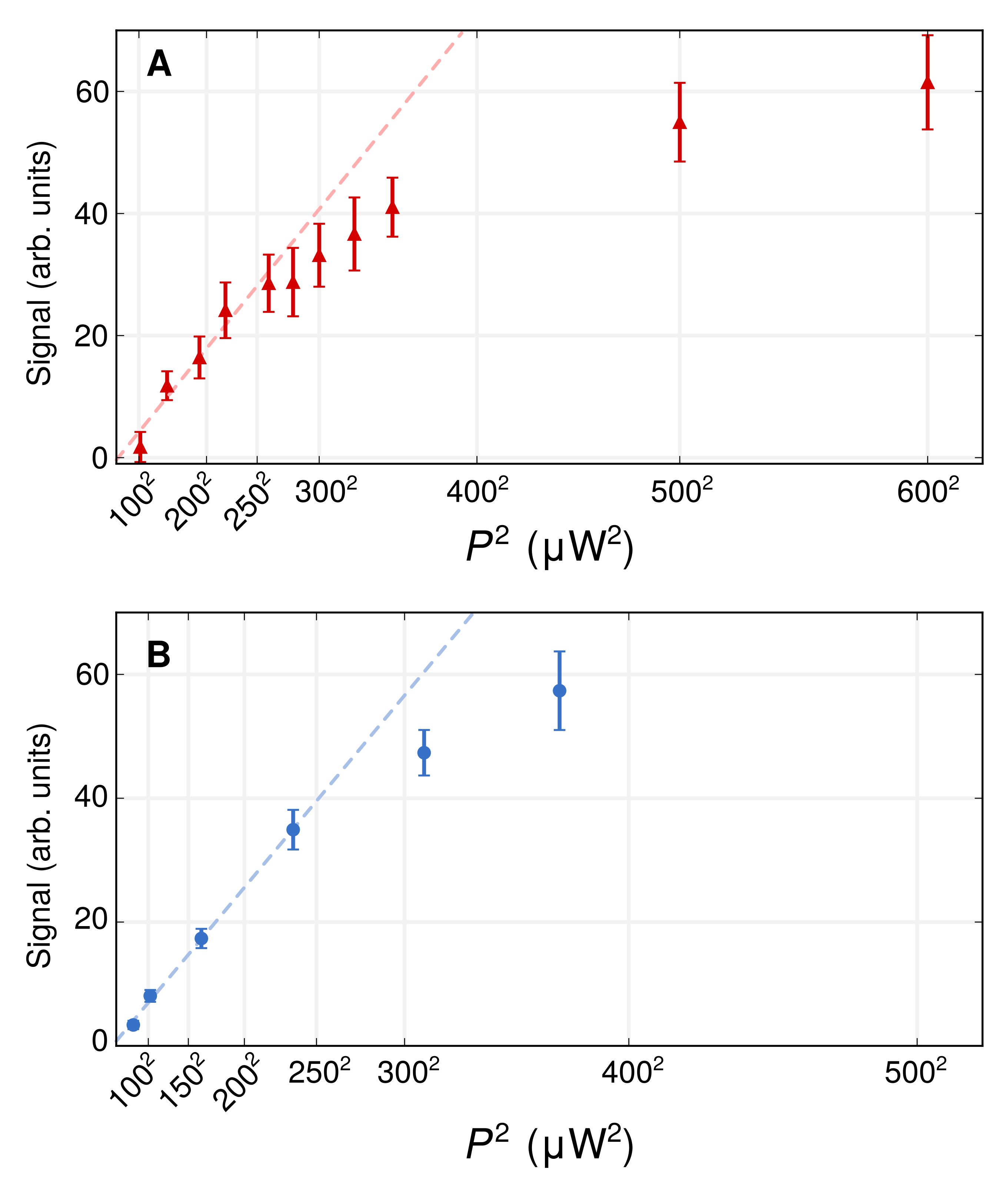}
\end{figure}
\noindent {\textbf{Fig. S2. Effect of probing power on Bragg signal}. Variation of the Bragg response as a function of probing power on each beam with $a$ = 0 $a_0$ for (\textbf{A}) the charge-mode (red triangles)  and (\textbf{B}) the spin-mode (blue circles). The horizontal axis is displayed in squared units for clarity. Dashed lines are fits within the linear-response regime.}

\section*{Supplementary Text}
\subsection{Controlling the number profile}

 The details of the number profile determine the ensemble dynamic structure factor (DSF), and thus the peak frequency and width of the measured Bragg spectrum. We measure the number profile in the 2D lattice by first using \textit{in situ}  PPCI, followed by the inverse Abel transform, which exploits the cylindrical symmetry of the array to obtain 3D densities. The measured number profile is an input into our calculations of the DSF and the expected Bragg spectrum. We ensure that the measured number profiles for all interaction strengths give a peak frequency of $10.0\pm0.15$ kHz for the calculated non-interacting Bragg spectrum. Thus, any shift in the peak frequency of the measured Bragg spectrum is due to the effect of interactions in 1D rather than systematic variation of the number profile. We tune the intensity of the compensation beams during the 3D lattice stage so that the anti-trapping potential of approximately $4\,E_r$ along each lattice axis for the non-interacting case reduces the density at the center of the lattice by the desired amount. Progressively less anti-trapping light is needed for stronger repulsive interactions to achieve an equivalent number profile. Figure S3 shows a number profile for a non-interacting gas and for $a$ = 500 $a_0$; the former required 4 $E_r$, while the latter 2.2 $E_r$, of compensation light. The compensation method is seen to be very effective.

\begin{figure}[hb]
\centering\includegraphics[width= 0.9\linewidth]{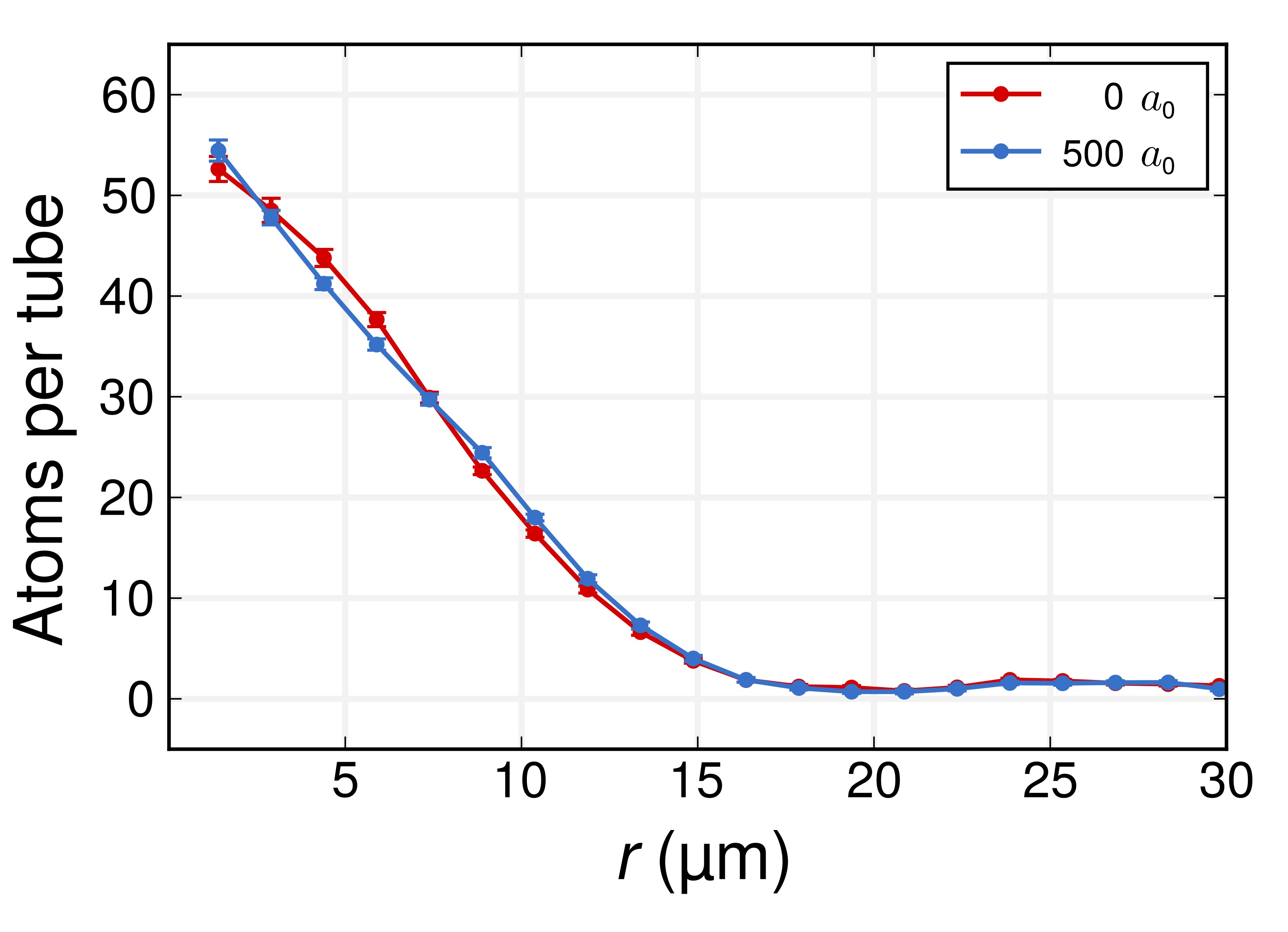}
\end{figure}
\noindent \textbf{Fig. S3. Number profile compensation.} Two representative number profiles showing the result of compensation with blue-detuned beams. The red data points show the number of atoms per tube for $a$ =0, compensated by 4 $E_r$ of green light per beam, while the blue points correspond to $a$ = 500 $a_0$, compensated by 2 $E_r$ per beam.
 
 \subsection{Effects of atom loss due to absorption of Bragg photons}
 
 Unlike in our measurements of the charge-mode Bragg spectrum, the absorption of Bragg photons during the spin-mode Bragg measurement is not negligible. Absorption events result in atom loss and occur at a rate $\propto I$, the intensity of the Bragg beams. We measured between $6-8\%$ loss during the spin-mode Bragg pulses, and for a fixed intensity, found no noticeable dependence on loss as a function of interaction. We determined that this atom loss does not have a significant impact on the measured spectra. Comparison between the charge-mode and spin-mode Bragg measurements for the non-interacting gas indicate that there is no detectable difference between the two spectra, as expected (see Fig. S4). 
 
\begin{figure}[ht!]
\centering\includegraphics[width= 0.9\linewidth]{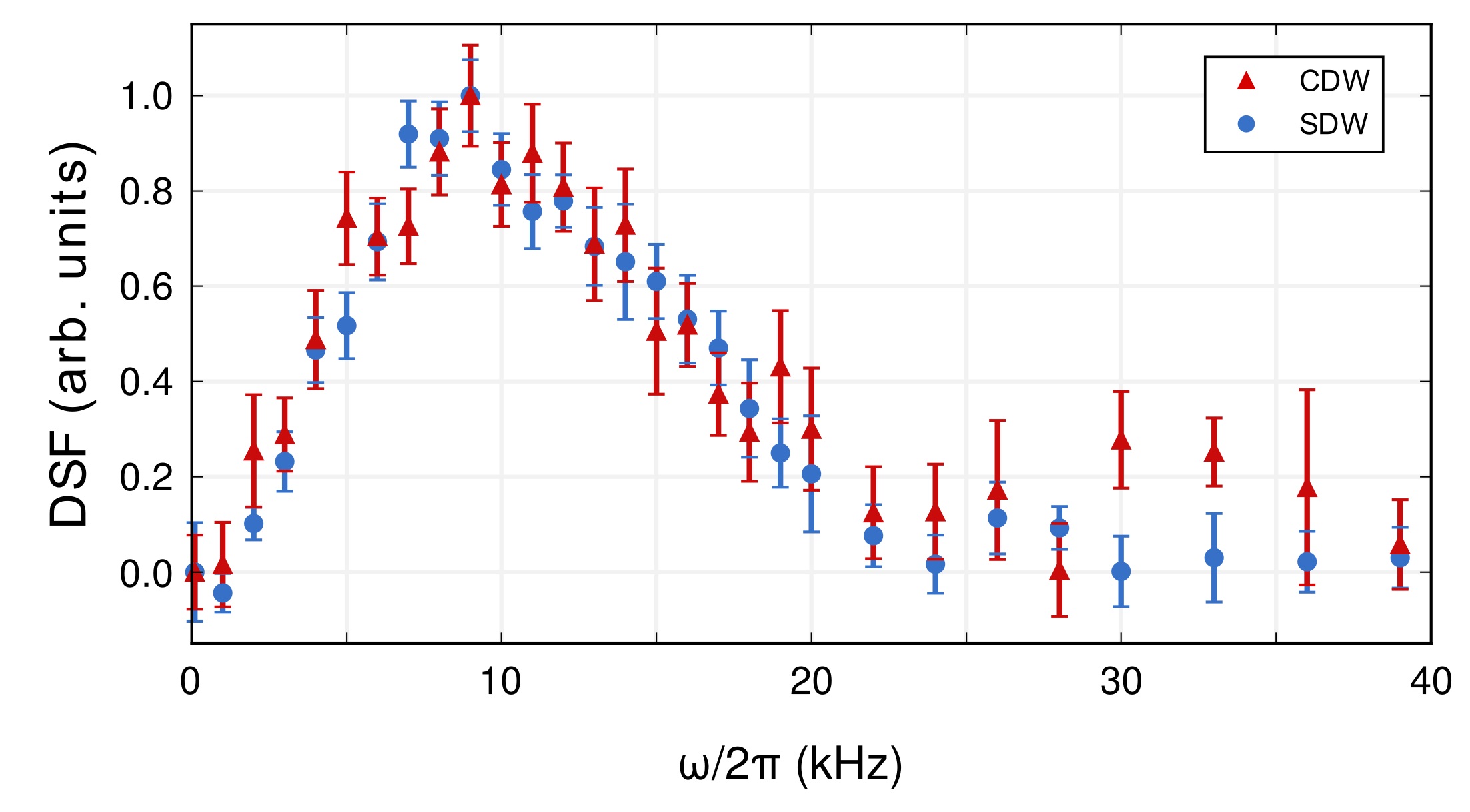}

\noindent \justifying{\textbf{Fig. S4. Non-interacting DSF comparison.}  Bragg spectra in the non-interacting case ($a$ = 0) between the charge-mode (red triangles) and the spin-mode (blue circles). The data are the same as that shown in Fig. 2, but for comparison purposes are displayed without offsetting the baseline. To within uncertainties, the two spectra are nearly indistinguishable along the whole excitation profile. We expect that the charge-mode signal at and above 30 kHz would decrease with further averaging.}
\end{figure}

In order to further confirm that our measured spin-mode Bragg spectra are unaffected by the amount of atom loss, we took further spin-mode measurements with even higher levels of loss (see Fig. S5). We performed these measurements at 0 and 500 $a_0$, and in each case probed the high-frequency region where thermal broadening might be expected. Increased loss of up to 15\% has no impact on the shape of the spectrum other than overall scaling, indicating that our results are not affected by atom loss. This could be due to the short timescale of the Bragg pulse.

\begin{figure*}[ht!]
\centering\includegraphics[width= 0.9\linewidth ]{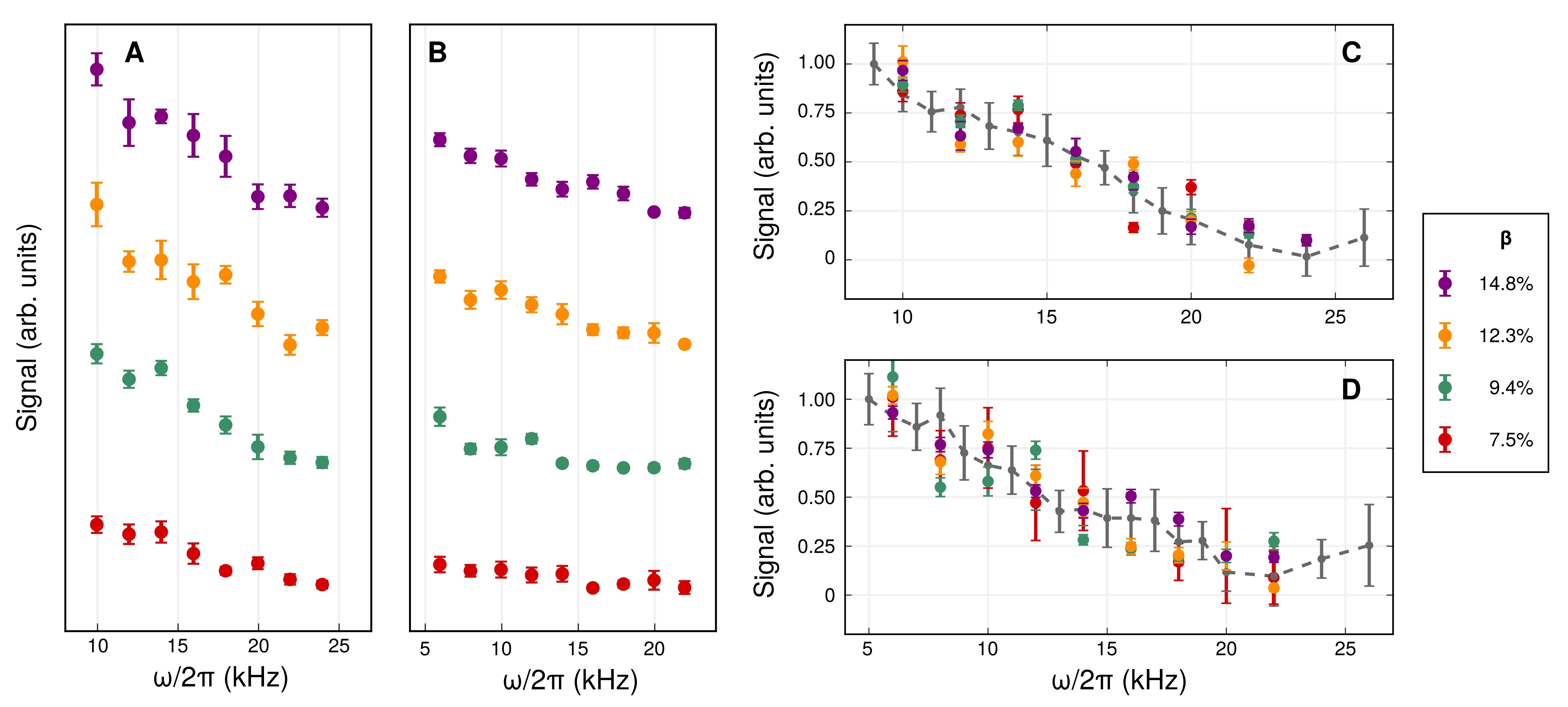}

 \noindent \justifying{\textbf{Fig. S5. Effects of atom loss on spin-mode Bragg spectrum.} Spin-mode Bragg spectra at 0 $a_0$ (\textbf{A}) and 500 $a_0$ (\textbf{B}) for high frequencies and different amounts of atom loss, $\beta$, resulting from different probing power (with fixed beam-waists). In each case a vertical displacement between data-sets with different values of $\beta$ is added for clarity. (\textbf{C},\textbf{D}) Same data-sets shown in (\textbf{A}) and (\textbf{B}) respectively, with each scaled according to maximum signal size. For comparison, the data in Fig. 2 in the main text are shown in gray for each case. Dashed lines are guides to the eye.} 
\end{figure*}
 
\subsection{Charge and spin density wave excitations}

If we consider a 1D two-component interacting Fermi gas consisting of an equal number ($N/2$) of spin-up ($\uparrow$) and spin-down ($\downarrow$) states, then the overall dynamic structure factor $S(q,\omega)$ (DSF) will have two independent components labeled by $S_{\uparrow\uparrow}$ and $S_{\uparrow\downarrow}$. Each component $S_{\sigma,\sigma'}$ is the Fourier transform of the density-density correlation function \cite{Pines_book}:

\begin{equation}
\begin{aligned}
S_{\sigma\sigma'}(q,\omega) &=\frac{1}{2\pi} \int {\rm d} z \int {\rm d}t\,  {\rm e}^{-i(q \cdot z-\omega t)}  \langle\rho_{\sigma}(x,t) \rho_{\sigma'}(0,0)  \rangle\\
&= \frac{1}{2\pi}  \int {\rm d}t\,  {\rm e}^{i\omega t}  \langle\rho_{\sigma}(q,t) \rho_{\sigma'}^{\dagger}(q,0)  \rangle,
\label{eq:g}
\end{aligned}
\end{equation}

\noindent where $\hat{\rho}_\sigma$ is the density operator for a spin state $\sigma$. At zero temperature, the momentum transfer to the atoms from the Bragg beams $P(q,\omega) \propto S(q,\omega)$, and using Fermi's golden rule we obtain:

\begin{equation}
P(q,\omega) \propto (R^2_{\uparrow}+R^2_{\downarrow})S_{\uparrow\uparrow}+2R_{\uparrow}R_{\downarrow}S_{\uparrow\downarrow},
\label{eq:pR}
\end{equation}

\noindent where $R_\sigma$ corresponds to the rate at which the atoms absorb a Bragg photon. For our experimental conditions, $R_\sigma\propto 1/\Delta_\sigma$, where $\Delta_\sigma$ is the relative detuning of the Bragg beam with respect to each spin state. Thus, we can finally write:

\begin{equation}
P(q,\omega) \propto \left(\frac{1}{\Delta^2_\uparrow}+\frac{1}{\Delta^2_\downarrow}\right)S_{\uparrow\uparrow}+\frac{2}{\Delta_\uparrow\Delta_\downarrow}S_{\uparrow\downarrow}.
\label{eq:pD}
\end{equation}

\subsection{Non-linear Luttinger Liquid}

The Hamiltonian of the 1D $\delta$-function interacting Fermi gas  is given by    
\begin{equation}
{\cal H}= -\frac{\hbar^2}{2 m} \sum_{i=1}^{N} \frac{\partial^2}{\partial z_i^2}+ 2c \sum_{1 \le i < j \leq N} \delta (z_i - z_j) -\mu N, \label{Ham}
\end{equation}
where $N$ is  the  total number of particles, including $N_{\uparrow}$ spin-up fermions and $N_{\downarrow}$ spin-down  fermions, and $\mu$ is the chemical potential. This is known as the Yang-Gaudin model and is Bethe ansatz solvable \cite{Yang1967,Gaudin1967}. In the above Hamiltonian (\ref{Ham}), the  coupling constant $c = g_1/2$.
 In the following theoretical analysis we first take $\hbar =1$,  and $2m =1$. 
 Later  we will recall  their   units for numerical simulation of charge and spin dynamical structure factors.
We define  a dimensionless interaction strength $\gamma = c / n$ with $n$ corresponding to the 1D density of the system	for our later  physical analysis.
In this supplemental material,  we consider  repulsive interactions, i.e.  $c >0$.

We obtain the density distribution functions $\rho_{c,s}$ via the following Bethe ansatz equations (BAE)  \cite{Yang1967,He2020}: 
\begin{eqnarray}
\rho_c (k) &=&\frac{1}{2 \pi}+ \int_{-\infty}^{\infty}a_1 (k-\lambda) \rho_s(\lambda) {\rm d}\lambda, \label{densityzerosH0-1}\\
\rho_s (\lambda)&=&\int_{-k_0}^{k_0}  s (\lambda-k) \rho_c (k) {\rm d}k, \label{densityzerosH0-2}
\end{eqnarray}
where  $s(\lambda)=1/(2c \cosh(\pi \lambda /c))$ and  
\begin{eqnarray}
a_n(k)=\frac{1}{2 \pi} \frac{n c}{ (nc)^2 /4+k^2}. \nonumber 
\end{eqnarray}
The zero-temperature charge dressed energy  $\varepsilon_c (k)$ and spin dressed energy $\phi_s (\lambda)$ are given by \cite{Guan2013,He2020} 
\begin{eqnarray}
\varepsilon_c (k) &=&k^2 -\mu + \int_{-\infty}^{\infty} a_1 (k-\lambda)\phi_s (\lambda) {\rm d} \lambda, \label{TBAzerosH0-1}\\
\phi_s (\lambda) &=&\int_{-k_0}^{k_0} s(\lambda-k) \varepsilon_c (k) {\rm d}k, \label{TBAzerosH0-2}
\end{eqnarray}
where $k_0$ is determined by the condition $n:\equiv N/L=\int_{-k_0}^{k_0}{\rho_c}(k)dk$. Using the  above equations, we may determine  the charge and spin velocities 
\begin{equation}
\label{vcvsdefination}
v_c=\frac{t_c}{2 \pi \rho_c(k_0)}, \quad  v_s=\frac{t_s}{2 \pi \rho_s(\lambda_0)}
\end{equation}
where $ t_c={\rm d}\varepsilon_c(k) /{\rm d} k |_{k=k_0}$ and  $ t_s={\rm d}\phi_s(\lambda) /{\rm d} \lambda |_{\lambda=\lambda_0}$. 

One can find distinct low-energy excitations \cite{He2020} in the charge and spin sectors.
As shown in Fig. S6, the charge sector has a particle-hole continuum spectrum, whose upper and lower bounds are given by
\begin{equation}
\omega_c(q)=v_c |q| \pm \frac{\hbar q^2}{2 m^*}+\cdots, \label{dispersion} 
\end{equation}
showing  a linear dispersion with quadratic curvature for arbitrarily strongly interacting fermions in the long-wave limit. 
In Eq.(\ref{dispersion}), $v_c$ is the charge velocity and $m^*$ is the effective mass, which can both be obtained from the BAE \cite{He2020}.
The low-energy spin excitation in the spin sector is also displayed in Fig. S6.
 It shows a typical two-spinon excitation spectrum, which are the elementary spin excitations for the Yang-Gaudin model   \cite{He2020}. 
 For a small momentum $q$,  the upper and lower bounds  are given by  
 \begin{equation}
\omega_{s+} (q)=v_s |q|-\frac{v_s {|q|}^3}{2k_s^2}   +\cdots, \qquad \omega_{s-} (q)=v_s |q|-\frac{2v_s {|q|}^3}{k_s^2}   +\cdots,\label{dispersion-s} 
\end{equation}
respectively, where $v_s$ is the spin velocity and $k_s$ is the characteristic spin wavevector.
The curvature leads to higher-order corrections in the nonlinear TLL \cite{Imambekov2009,Imambekov2012}. 
We observe that the upper boundary of the two-spinon excitation spectrum is determined by two spinons with equal quasi-momentum, whereas the lower boundary is determined by two separated spinons. 
 
\sloppy In accordance with bosonization theory, one introduces  charge and spin bosonic fields 
 $\phi_{c,s}=\left( \phi_{\uparrow}\pm \phi_{\downarrow}\right)/\sqrt{2}$ and $\Pi_{c,s}=\left( \Pi_{\uparrow}\pm \Pi_{\downarrow}\right)/\sqrt{2}$. Here the fields $ \phi_{\nu}$ and their canonically conjugate momenta $\Pi_{\nu}$ obey the standard bosonic commutation relations, i.e. 
$\left[ \phi_{\nu},\Pi_{\mu} \right]=\mathrm{i} \delta_{\nu\mu}\delta(x-y)$ with $\nu,\mu=c,s$.  
For small momentum $q\ll k_F$,  the 1D spin-$1/2$ repulsive Fermi gas can be described by an effective Hamiltonian \cite{Imambekov2012,Giamarchi_Book,Haldane1981}
\begin{equation}
H=H_c+H_{s}+\frac{2g_1}{(2\pi \alpha)^2}\int dz \cos (\sqrt{8}\phi_{s}). \label{effective}
\end{equation}

\noindent The parameter $\alpha$ is a short-range cutoff.  
The  first two terms  describe the conventional linear TLL, i.e. the low-energy excitations of the 1D interacting Fermi gas separates into charge and spin collective bosonic modes, with
\begin{equation}
H_{\nu}=\int dx \left(\frac{\pi v_{\nu} K_{\nu}}{2} \Pi^2_\nu+\frac{v_{\nu}}{2\pi K_{\nu} }\left(\partial _z \phi_{\nu} \right)^2\right).\label{Linear-TLL}
\end{equation}
This effective Hamiltonian characterizes the long-distance asymptotic decay of correlations in the 1D Fermi gas, i.e. the linear TLL behavior.
The coefficients  for different processes are given phenomenologically in Ref. \cite{Giamarchi_Book}.  
The coefficient $v_c/K_c$ is the energy cost for changing the particle density,  while $v_{s}/ K_{s}$ 
determines the energy for creating a nonzero spin polarization. 
The last term in the effective Hamiltonian (\ref{effective})  characterizes the $2k_F$ back-scattering  process, where
$g_1$ is the coupling constant of this marginally irrelevant back-scattering operator (in the RG sense). This back-scattering term only affects the spin sector, and as we will show, it has a quite significant effect on the spin DSF. In the above TLL description, the velocities $v_{c,s}$ and Luttinger liquid parameters $K_{c,s}$ can be calculated using the Bethe ansatz and thermodynamic equations \cite{Guan2013}, via
\begin{eqnarray}
\chi=\frac{K_s}{2\pi v_s},\quad\kappa = \frac{2K_c}{\pi v_c}, \label{Ks-Kc}
\end{eqnarray}
where $ \chi $ and $\kappa$ are the spin susceptibility and charge compressibility, respectively. 

\begin{figure}[hb!]
	\centering
	{\includegraphics[width= 0.9\linewidth]{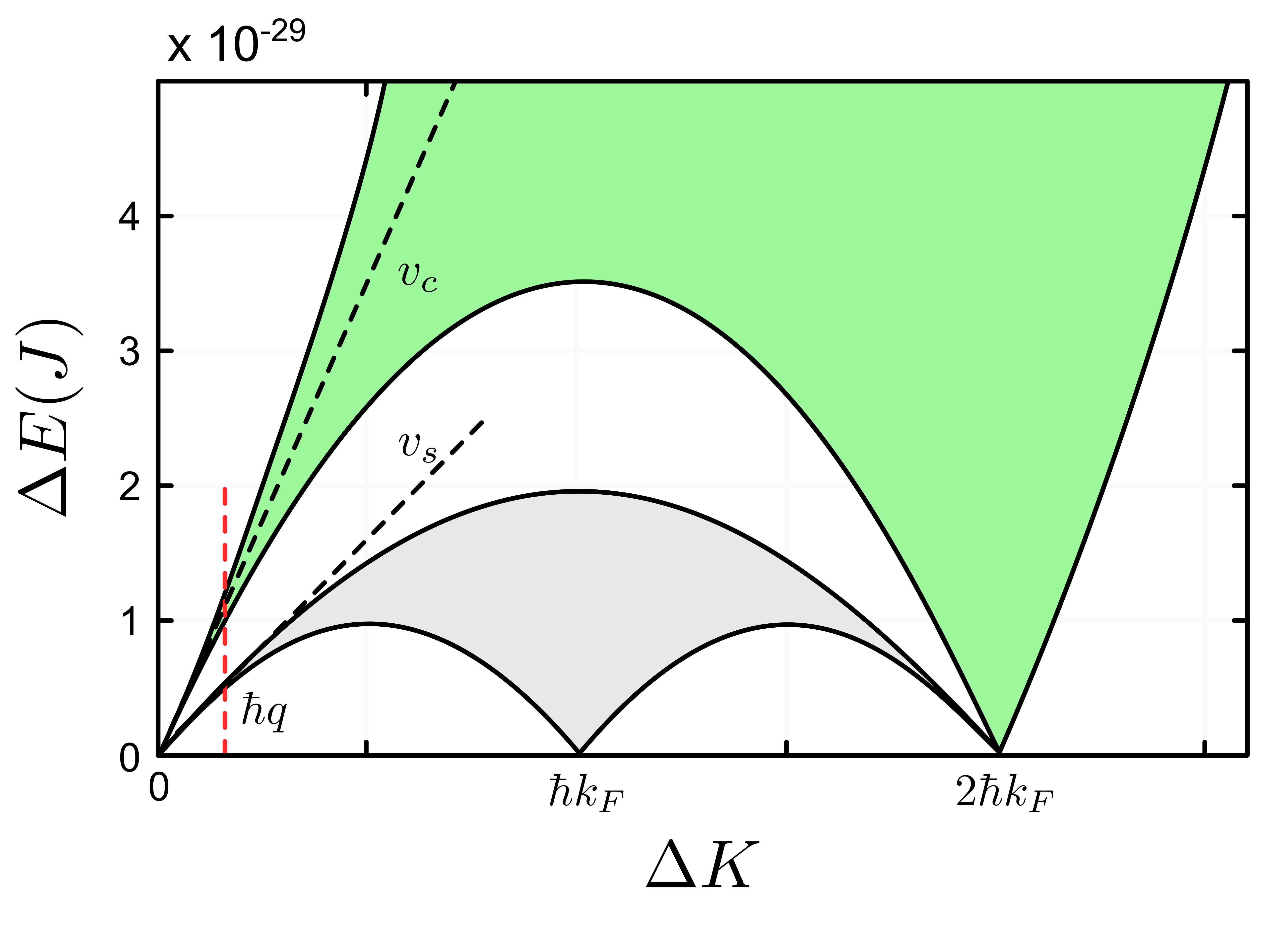}} 	
	\label{spectrurm}

\noindent \justifying{\textbf{Fig. S6. Excitation Spectrum for Yang-Gaudin Model.} Exact particle-hole (green) and two-spinon (gray) excitation spectra for a repulsive Fermi gas with periodic boundary condtions at $\gamma=c/n = 5.03$ with the Fermi surface $k_F=n\pi$, where density $n=N/L=3 \times 10^6 \;(\rm 1/m)$, $\Delta E=\hbar \omega$. The black dashed lines in the charge and spin spectra correspond to the  charge velocity $v_c$ and spin velocity $v_s$, respectively.  Here the red dashed line shows the excited momentum imparted by the Bragg beams in our experiment, which is set as $\Delta K= \hbar q$,  $q =1.47\;\mu m^{-1}$  for both charge and spin DSFs.}
\end{figure}

\subsection{Charge and Spin DSFs}

Although we have exact solutions of the model (Eq.~(\ref{Ham})), the charge and spin dynamic structure factors for a repulsive Fermi gas have yet to be analytically calculated. 
This is a long standing theoretical challenge. In the linear TLL theory at finite temperature, the charge  DSF is a $\delta$-function at $\omega = v_c q$ for $q\ll k_F$, i.e.
\begin{eqnarray}
S_C(q,\omega)=\frac{qK_c }{2\pi^2}\delta(w-v_c q). \label{charge-DSF-delta}
\end{eqnarray}

\noindent As the leading correction to the linear dispersion for charge is quadratic in $q$, see Eq.~(\ref{dispersion}), the charge DSF of a Fermi gas with finite repulsive interaction can be well approximated by that of a non-interacting ideal Fermi gas \cite{Cherny2006,Pereira2010}:
\begin{equation}
S(q,\omega) = \frac{{\rm Im}\chi(q,\omega, k_F,T,N)}{\pi (1-e^{-\beta\hbar \omega})}, \label{SC}
\end{equation}
valid for $T\ll T_F=mv_F^2/2$ and $q \ll k_F$. In the above equation  $\chi$ is the dynamic polarizability, the imaginary part of which is
\begin{eqnarray}
\label{finiteTDSF}
{\rm Im} \chi (q, \omega,k_{F},T,N)=\frac{N  m^*}{2 \hbar^2 q k_F} \pi (n_{q_{-}}-n_{q_{+}}),
\end{eqnarray}
where 
\begin{eqnarray}
\label{fermidistribution}
q_{\pm}=\frac{\omega m^*}{\hbar q}\pm \frac{q}{2},\quad n_k=\frac{1}{{\rm e}^{\beta(\varepsilon -\mu)}+1},\quad \varepsilon=\frac{\hbar^2 k^2}{2 m^*}.
\end{eqnarray}

\noindent The interaction effect can be accounted for by replacing $k_F$ with $k_c=m^*v_c/\hbar$, as was done in previous work \cite{Ernie2018}. Incorporating broadening due to finite $q$ is necessary to model our charge-mode Bragg spectra. Fig. S7 shows a comparison of measured spectra and theoretical spectra, calculated either using the linear DSF (\ref{charge-DSF-delta}) or the free-fermion DSF (\ref{SC}).

\begin{figure}[hb!]
\label{charge-DSF}
\centering\includegraphics[width= 0.9\linewidth]{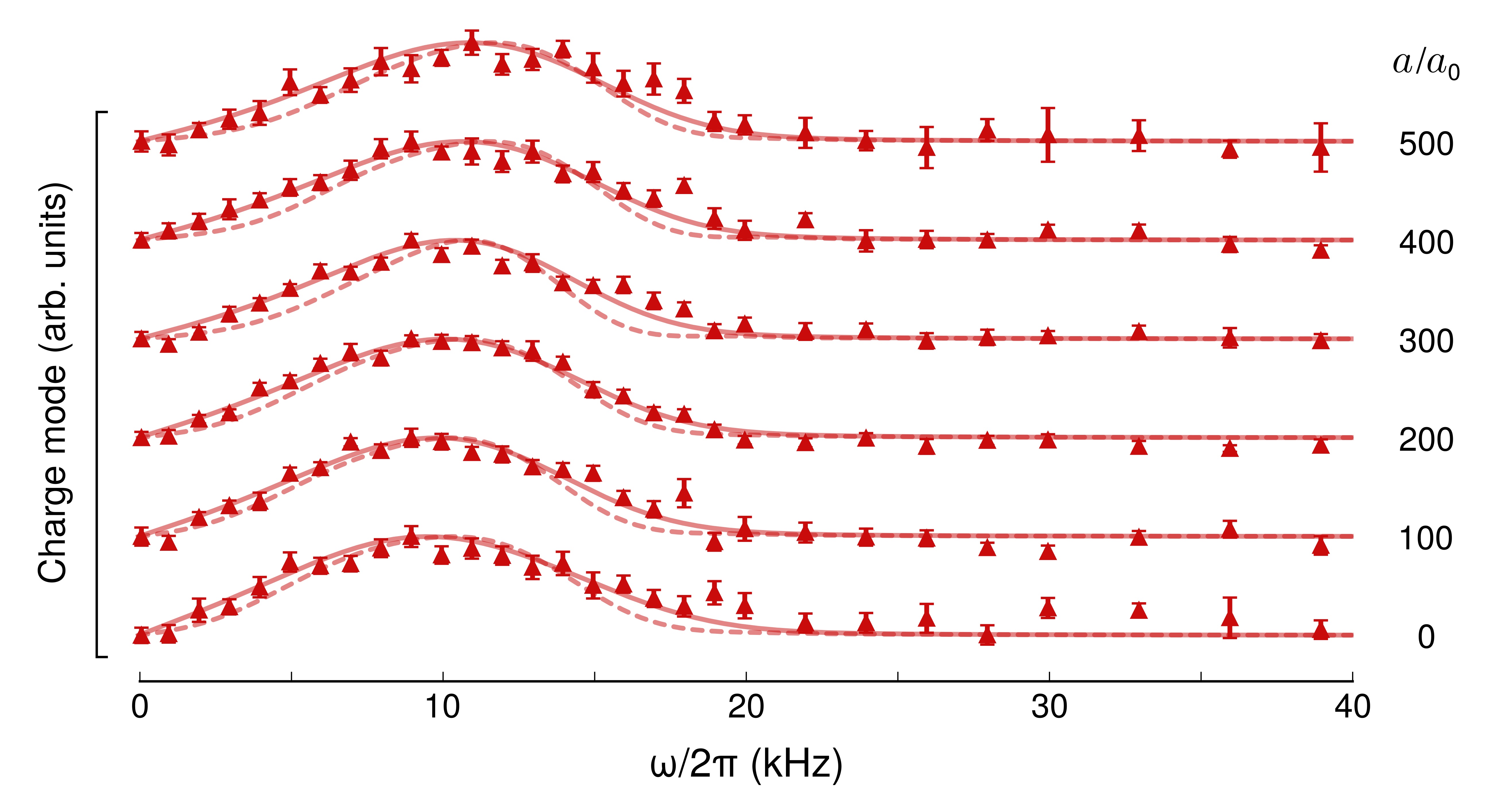}

\noindent \justifying{\textbf{Fig. S7. Comparison of linear and nonlinear models of $S_C$.} Measured (symbols) and calculated (lines) charge-mode Bragg spectra for interactions in the range from 0 to 500 $a_0$. Solid lines calculated using free-fermion DSF (\ref{SC}) at $T=250$ nK, which includes effects of band-curvature. Dashed lines calculated using linear DSF (\ref{charge-DSF-delta}).}
\end{figure}

For the spin DSF, we do not have a similar approximation in terms of non-interacting spinless fermions. At zero temperature, the spin DSF is a $\delta$-function peak at $\omega = v_s q$.
This will be broadened by band curvature and other irrelevant interactions (in the RG sense) that may lead to spin-charge coupling \cite{Pereira2010}. To calculate the contribution from these high order terms to the spin DSF remains extremely challenging, but they should scale as $q^3$. Hence,
for $q\ll k_F$ and a temperature $T\sim v_s q \ll T_F=mv_F^2/2$, the elementary spin excitations are essentially captured by the effective Hamiltonian (\ref{effective}) which includes only the back-scattering term in addition to the linear TLL Hamiltonian. 
%

%


 At zero temperature, back-scattering contributes to the spin DSF as a $\delta$-function, i.e.
\begin{eqnarray}
S_S(q,\omega)=\frac{qK_s }{2\pi^2}\delta(w-v_s q).\label{spin-DSF-delta}
\end{eqnarray}

\noindent At finite temperature, however, such a $\delta$-function peak is broadened.
 Therefore, the elementary excitations in the spin sector are bosons with finite lifetimes and the propagator of the dressed spin bosons is given by \cite{Pereira2010}
\begin{equation}
\widetilde{S}(q,i\omega)=\frac{1}{4\pi}\frac{q}{i\omega-v_sq-\sum(q,i\omega,T)},\label{Spin-propagator}
\end{equation}
where $\sum(q,i\omega,T)$ is the self-energy of spin bosons,
whose exact expression is very difficult to calculate. 
%
Here, based on Fermi's golden rule, we apply an approximate method to calculate the propagator in Eq.~\ref{Spin-propagator} \cite{Balents2001}.
We assume that the real part of the retarded self-energy is zero (thus the mass shell of the spin boson is still $w=v_{s}q$), while the imaginary part is given by 
\begin{equation}
\mathrm{Im} \sum\nolimits_q^{ret}=-\frac{1}{\tau_s(T)}=-\frac{\pi}{2}[g(T)]^2 k_\mathrm{B}T.\label{spin-DSF-d}
\end{equation}
where 
\begin{eqnarray}
g(T)\approx\frac{g}{1+g\ln(T_F/T)},
\end{eqnarray}
is the renormalized coupling constant at finite termperature $T$ and we use $g= g_1/(\pi v_s)$.
We thus obtain the retarded spin-spin correlation function 
\begin{eqnarray}
\chi^{ret}(q,\omega)=-\frac{2K_s}{\pi}\frac{q}{\omega-v_sq+i/\tau_s(T)}\,,
\label{ret-spin-spin}
\end{eqnarray}
the imaginary part of which leads to the spin DSF.

By comparing the calculations to our measured data, we determine that including the back-scattering term is necessary to model the spin Bragg spectra, particularly for large interactions (see Fig. S8). The linear TLL model fails to reproduce the observed high-frequency tails of the spin-mode Bragg spectra. Further deviations from theory at high frequency may be due to unaccounted-for corrections of order $q^3$.
 
\begin{figure}[ht!]
\label{Spin-DSF}
\centering\includegraphics[width= 0.9\linewidth]{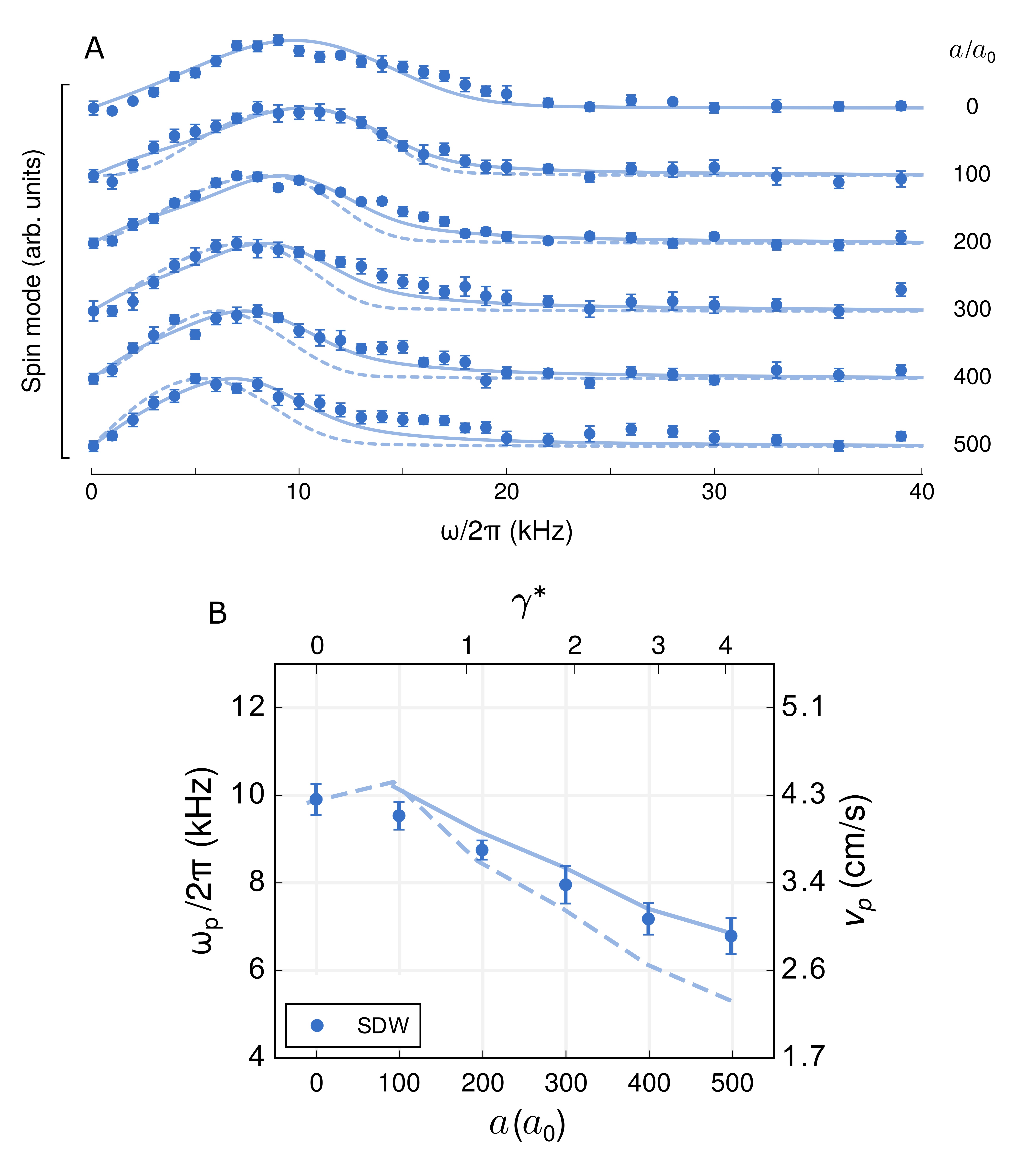}

\noindent \justifying{\textbf{Fig. S8. Comparison of linear and nonlinear models of $S_S$.} (\textbf{A}) Measured (symbols) and calculated (lines) spin-mode Bragg spectra for interactions in the range from 0 to 500 $a_0$. Solid lines incorporate effects of back-scattering (\ref{ret-spin-spin}) at $T=250$ nK. Dashed lines calculated using linear DSF (\ref{spin-DSF-delta}). (\textbf{B}) Peak frequencies of theoretical and experimental spin DSF. We observe that for weak interactions, the theoretical peaks are larger than experimental ones, perhaps due to an inability to accurately model the contributions from  the band curvature.}
\end{figure}

\subsection{Computing the Ensemble DSF}

In our experiment, we prepare the ultracold $^6$Li gas in a 2D optical lattice, in which quasi-1D tubes are formed along the $x$ direction.
We apply the local density approximation (LDA) to treat physical quantities of  the harmonically trapped 1D ultracold atomic system, where the density distribution satisfies  the conditions 
\begin{equation}
\label{LDAdensity}
\left\{
\begin{array}{lr}
\mu (n_{\mathrm{1D}}(z))=\mu_0-V(z),\quad z\leq R_F,\\
n_{\mathrm{1D}}(z)=0,\quad \quad \quad \quad \quad \quad \; z>R_F.
\end{array}
\right.
\end{equation}
Here, $R_F$ is Thomas-Fermi radius, $\mu (n_{1D}(z))$ is the effective local chemical potential and $V(z)=m \omega^2 z^2/2$ is the trapping potential.
In this case, the total number in the 1D tube is given by 
\begin{equation}
\label{Nof1D}
N(\mu_0)=\int n_{\mathrm{1D}}(\mu, z) {\rm d}z,
\end{equation}
where the 1D density $ n_{\mathrm{1D}}(\mu, z)$ can be exactly calculated in terms of quasi-momentum density functions using the homogenous chemical potential $\mu_0$ and the so-called Thermodynamic Bethe Ansatz (TBA) equations \cite{Takahashi1971}.

For a given total number $N$ in a 1D tube, one  may obtain the density distribution along the tube direction by solving density functions and choosing a proper central chemical potential $\mu_0$ in (\ref{LDAdensity}) to enforce Eq.~(\ref{Nof1D}). 
Furthermore,  one may calculate the average DSF for a  1D tube via 
\begin{equation}
\label{1DDSFLDA}
S_{\mathrm{1D}}(q,\omega,N)=\int S_0(q,\omega,n_{\mathrm{1D}}(z)) n_{\mathrm{1D}}(z) {\rm d}z.
\end{equation}
Figure S9 shows the calculated Bragg signal for a harmonically confined gas and a homogeneous one, at zero temperature and at 250 nK. The total DSF of a 3D sample is given by 
\begin{equation}
\label{3DDSF}
S_{\mathrm{3D}}=\sum_i S_{\mathrm{1D}}(q,\omega,N(r_i)) N_{t}(r_i),
\end{equation}
where $N(r_i)$ is the average particle number of each 1D tube at a radial distance $r_i$ from the center of the array, $i$ is the radial lattice index and $N_{t}(r_i)$ is the number of tubes at a distance $r_i$ from the center. Figure S10 shows the contributions to the calculated Bragg signal from different shells of tubes at different distances from the center of the lattice for a measured number profile. The relative size of each contribution depends on $N(r_i)$ (which is peaked at the center of the trap), and $N_t(r_i)$, which increases linearly with $r_i$. In our samples, the peak contribution is from tubes with approximately 30 atoms.

\begin{figure}[ht!]
\label{SingleTubeLDA}
\centering\includegraphics[width= 0.9\linewidth]{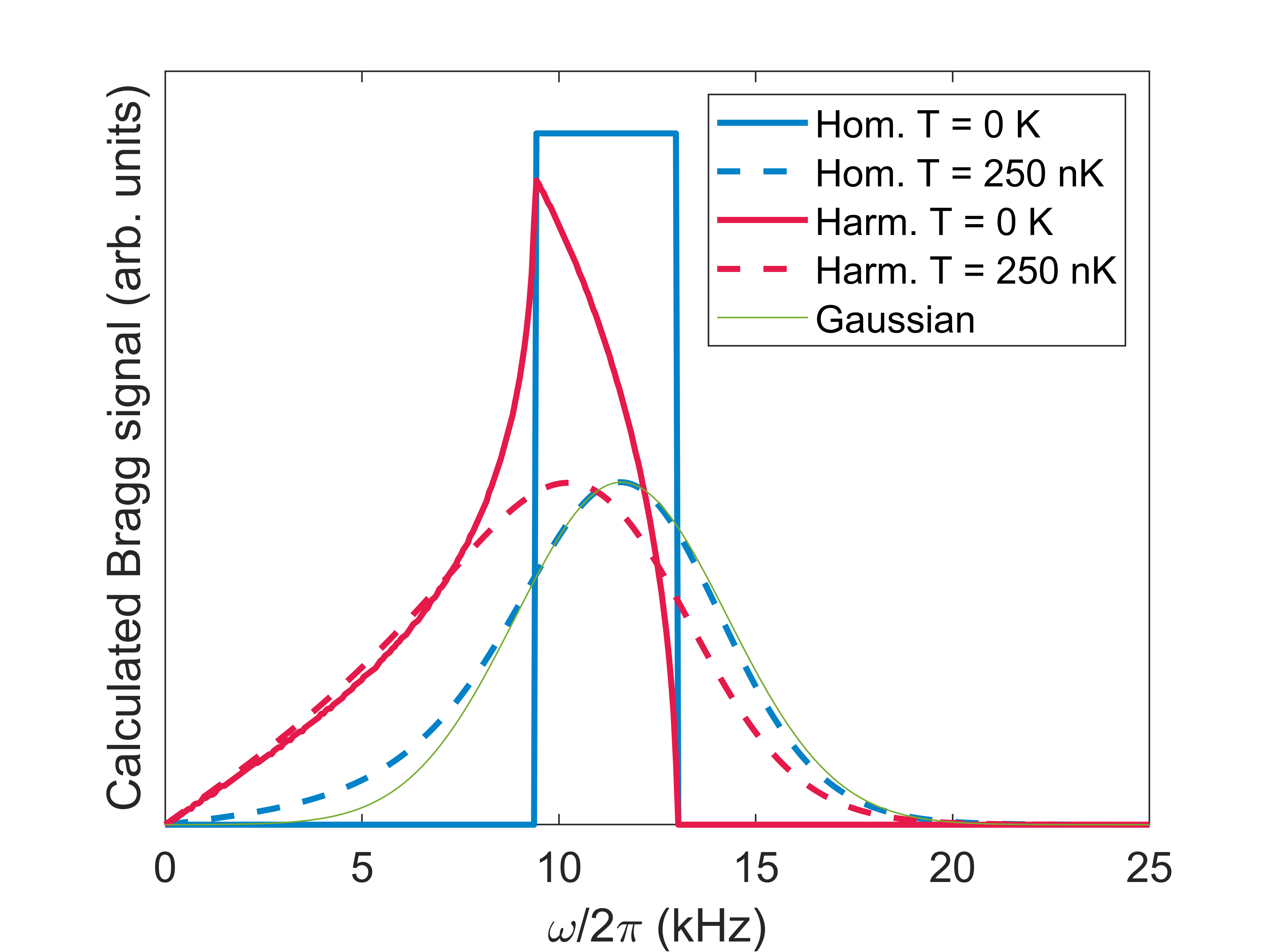}

\noindent \justifying{\textbf{Fig. S9. Comparison of Bragg signal for homogeneous and harmonically confined gases.} Calculated Bragg signal for a single tube containing 26 non-interacting atoms, assuming harmonic confinement ($\omega_z = 2\pi \times 1.34$ kHz), for $T=0$ K (red, solid) and 250 nK (red, dashed). Calculated Bragg signal for a homogeneous gas with same Fermi temperature as harmonic case, for $T=0$ K (blue, solid) and 250 nK (blue, dashed). Pulse-time broadening effects neglected for clarity. Gaussian fit to finite-$T$ DSF for homogeneous gas shown in green (full-width at half-maximum of $2\pi\times$6.3 kHz).}
\end{figure}

The momentum transferred due to Bragg scattering is further broadened due to the finite duration of the Bragg pulse. We account for this broadening using  \cite{Zou2010}

\begin{equation}
\label{duration}
P(q,\omega) \propto \frac{1}{\pi \sigma}\int_{-\infty}^{\infty} { S_{\mathrm{3D}}(q,\omega') {\rm sinc}^2 \left[\frac{\omega-\omega'}{\sigma}\right]\rm d}\omega',
\end{equation}

\noindent where ${\rm sinc}(x)=\sin(x)/x$ and the energy resolution $\sigma=2/\tau_{\rm B}$ is set by the experimental Bragg pulse duration time $\tau_{\rm B}=200\,\mu$s.

\begin{figure}[ht!]
\centering\includegraphics[width= 0.9\linewidth]{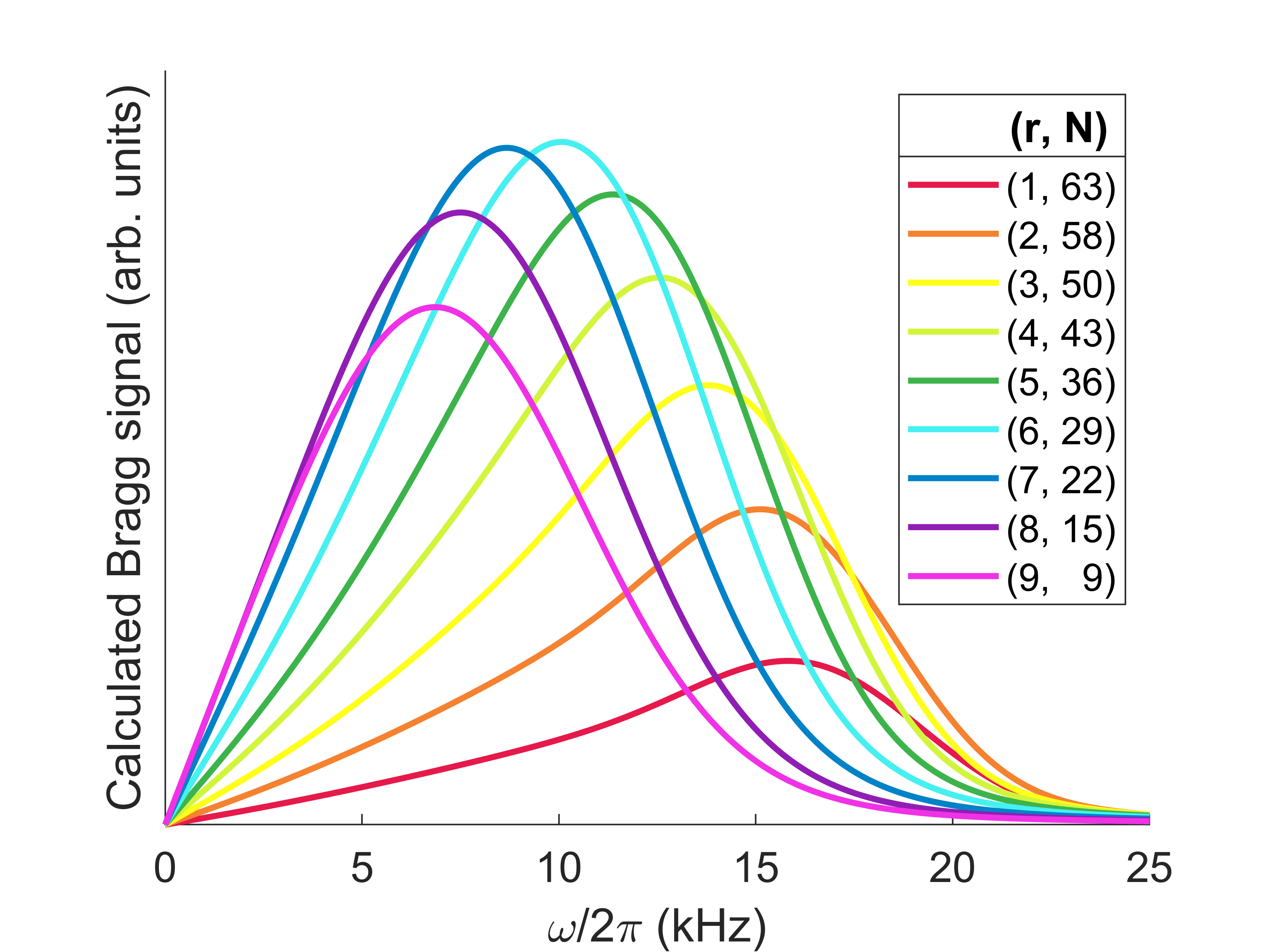}

\noindent \justifying{\textbf{Fig. S10. Contributions to Bragg signal from tube ensemble.} Calculated Bragg signal for collection tubes located a radial distance $r$ (arb. units) from center of trap, which have an average tube occupancy of $N$ atoms. Interactions assumed to be zero, and number profile is taken from measurement of $S_C(q,\omega)$ for $0\,a_0$.}
\end{figure}



\subsection{Effect of back-scattering on out-coupled width}

 The broadening of the spin DSF in $\omega$-space due to back-scattering explains the increase in broadening of the out-coupled atom packet in position-space after time-of-flight expansion (see Fig. 4 of the main text). Here we include a description of the method used to compute the out-coupled widths for the spin-mode for the probed range of interaction strengths.
 
 The spin-mode DSF is a Lorentzian, with a full-width at half-maximum (FWHM) given by $2/\tau_s(T)$ (\ref{ret-spin-spin}). The non-interacting DSF for a local region of uniform density at finite $T$ is well approximated by a Gaussian (see Fig. S9). We combine the FWHM of these two DSFs according to the convolution of a Lorentzian and a Gaussian (a Voigt profile). By dividing the convolved FWHM by the original Gaussian FWHM, we obtain a relative broadening, which we then scale by the width of the out-coupled atoms at 0 $a_0$ to obtain the theory line shown in Fig. 4.

\end{document}